\documentstyle[12pt,epsfig]{article}

\newcommand{\nn}{\nonumber}
\newcommand{\beq}{\begin{equation}}
\newcommand{\eeq}{\end{equation}}
\newcommand{\ba}{\begin{eqnarray}}
\newcommand{\ea}{\nonumber \end{eqnarray}}
\newcommand{\be}{\begin{eqnarray}}
\newcommand{\ee}{\nonumber \end{eqnarray}}

\newcommand{\bs}{\begin{slide}}
\newcommand{\es}{\end{slide}}
\newcommand{\bc}{\begin{center}}
\newcommand{\ec}{\end{center}}
\newcommand{\bi}{\begin{enumerate}}
\newcommand{\ei}{\end{enumerate}}

\def\fun#1#2{\lower3.6pt\vbox{\baselineskip0pt\lineskip.9pt
\ialign{$\mathsurround=0pt#1\hfil##\hfil$\crcr#2\crcr\sim\crcr}}}

\begin{document}

\title{Sector of the $2^{++}$ mesons: observation of the tensor
glueball}
\author{V.V. Anisovich$^a$, M.A. Matveev$^a$, J. Nyiri$^b$ and
A.V.~Sarantsev$^{a,c}$\\
$^a$ Petersburg Nuclear Physics Institute, Gatchina 188300, Russia\\
$^b$ KFKI Research Institute for Particle \\and Nuclear Physics,
Budapest, Hungary\\
$^c$ HISKP, Universit\"at Bonn, D-53115 Germany}

\date{17.03.05}
\maketitle

\begin{abstract}

Data of the Crystal Barrel and L3 collaborations clarified
essentially the situation in the $2^{++}$ sector in the mass region
up to 2400 MeV, demonstrating the linearity of $(n,M^2)$
trajectories, where $n$ is the radial quantum number of a
quark-antiquark state with mass $M$. We discuss these data and show
that there exists a superfluous state for the $(n,M^2)$
trajectories: a broad resonance $f_2(2000)$. We pay special
attention to the reactions $p\bar p\to\pi\pi,\eta\eta,\eta\eta'$ in
the mass region 1990--2400~MeV where, together with $f_2(2000)$,
four relatively narrow resonances are seen: $f_2(1920)$,
$f_2(2020)$, $f_2(2240)$, $f_2(2300)$. We analyse the branching
ratios of all these resonances and show that only the decay
couplings of the broad state $f_2(2000)\to\pi^0\pi^0,\eta\eta,
\eta\eta'$ satisfy relations inherent in the glueball decay.
\end{abstract}


\section{Introduction}

A broad isoscalar-tensor resonance in the region of 2000 MeV is
seen in various reactions \cite{PDG}. Recent measurements give: \\
$M=2010\pm25\,$MeV, $\Gamma=495\pm35\,$MeV in
$p\bar p\to\pi^0\pi^0, \eta\eta, \eta\eta'$ \cite{Ani},\\
$M=1980\pm20\,$MeV, $\Gamma=520\pm50\,$MeV in $pp\to pp\pi\pi\pi\pi$
\cite{Bar},\\
$M=2050\pm30\,$MeV, $\Gamma=570\pm70\,$MeV in $\pi^-p\to\phi\phi n$
\cite{LL};\\
following them, we denote the broad resonance as $f_2(2000)$.

The large width of $f_2(2000)$ arouses the suspicion that this
state is a tensor glueball. Such an opinion was expressed lately
in different publications.

In \cite{book}, Chapter 5.4, it is said that the very broad
isoscalar $2^{++}$ state observed in the region $\sim 2000$ MeV with
a width of the order of $400-500$ MeV \cite{Bar} could well be the
trace of a tensor glueball lying on the Pomeron trajectory.

Another argument comes from the analysis of the mass shifts of the
$q\bar q$ tensor mesons (\cite{L3}, Section 12). It is stated here
that the mass shift between $f_2(1560)$ and $a_2(1700)$ can not be
explained by the mixing of non-strange and strange components in
the isoscalar sector: in such a mixing the average mass squared
does not change and we should find $f_2(1750)$ at a much higher
mass. Instead, we observe a shift down in masses of both isoscalar
states. Such a phenomenon can be an indication for the presence of
a tensor glueball in the mass region 1800-2000 MeV.

In \cite{LL}, the following argument is presented: a significant
violation of the OZI-rule in the production of tensor mesons with
dominant $s\bar s$ components (reactions $\pi^-p\to f_2(2120)n$,
$f_2(2340)n$, $f_2(2410)n \to\phi\phi n$ \cite{Etk}) is due to the
presence of a broad glueball state $f_2(2000)$ in this region,
resulting in a noticeable admixture of the glueball component in
$f_2(2120)$, $f_2(2340)$, $f_2(2410)$.

The possibility that the broad resonance $f_2(2000)$ could be a
glueball is discussed also in \cite{Bugg}, Section 10. Here,
however, a problem in the identification of $f_2(2000)$ as the
tensor glueball is stressed. As it is written in Subsection 10.6, of \cite{Bugg}
the prediction for the branching fraction of the $2^+$ glueball is
large if the width is taken to be the 500 MeV fitted to $f_2(1950)$.
Observed decays to $\sigma\sigma$ and $f_2(1270)\sigma$ account for
$(10\pm 0.7\pm 3.6)\cdot 10^{-4}$ of $J/\Psi$ radiative decays and
for a further $(7\pm 1\pm 2)\cdot 10^{-4}$ in $K^*\bar K^*$ decays.
If one assumes flavour-blindness for vector-vector final states, the
vector-vector contribution increases to $(16\pm 2\pm 4.5)\cdot
10^{-4}$. The total $2.6\cdot 10^{-3}$ is still less by a factor 9
than predicted for a glueball; in \cite{Bugg} this is considered as
a problem in identifying $f_2(1950)$ with the $2^+$ glueball.

In \cite{glueball2} it was emphasised that the $f_2(2000)$ being
superfluous for $q\bar q$ systematics can be considered as the
lowest tensor glueball. A recent re-analysis of the $\phi\phi$
spectra \cite{LL} in the reaction $\pi^-p\to\phi\phi n$
\cite{Etk}, the study of the processes $\gamma\gamma\to
\pi^+\pi^-\pi^0$ \cite{L3-3pi}, $\gamma\gamma\to K_SK_S$ \cite{L3}
and the analysis of the $p\bar p$ annihilation in flight $p\bar
p\to\pi\pi, \eta\eta, \eta\eta'$ \cite{Ani} clarified essentially
the status of the ($J^{PC}=2^{++}$)-mesons. This allows us to
place the $f_2$ mesons reliably on the $(n,M^2)$-trajectories
\cite{glueball2}, where $n$ is the radial quantum number of the
$q\bar q$-state. In the present review we discuss the data for
$\gamma\gamma\to \pi^+\pi^-\pi^0$ \cite{L3-3pi}, $\gamma\gamma\to
K_SK_S$ \cite{L3} and $p\bar p\to\pi\pi, \eta\eta, \eta\eta'$
\cite{Ani} in Section 2.

In \cite{syst} (see also \cite{book,ufn04}), the known $q\bar
q$-mesons consisting of light quarks ($q=u,d,s$) are placed on the
$(n,M^2)$ trajectories. Trajectories for mesons with various
quantum numbers turn out to be linear with a good accuracy. In
Section 3 we give a systematisation of tensor mesons, $f_2$ and
$a_2$, on the $(n,M^2)$ planes.

The quark states with ($I=0$, $J^{PC}=2^{++}$) are determined by
two flavour components $n\bar n=(u\bar u+d\bar d)/\sqrt 2$ and $s\bar s$
for which two states $^{2S+1}L_J=\,^3P_2,\,^3F_2$ are possible.
Consequently, we have four trajectories on the $(n,M^2)$ plane.
Generally speaking, the $f_2$-states are mixtures of both the flavour
components and the $L=1,3$ waves. The real situation is, however, such
that the lowest trajectory [$f_2(1275)$, $f_2(1580)$, $f_2(1920)$,
$f_2(2240)$] consists of mesons with dominant $^3P_2n\bar n$
components, while the trajectory $[f_2(1525)$, $f_2(1755)$, $f_2(2120)$,
$f_2(2410)]$ contains mesons with predominantly $^3P_2s\bar s$
components. The $F$-trajectories are presently represented by
three resonances [$f_2(2020)$, $f_2(2300)$] and [$f_2(2340)$] with
the corresponding dominant $^3F_2n\bar{n}$ and $^3F_2s\bar s$
states. In \cite{glueball2}, it is shown that the broad resonance
$f_2(2000)$ is not part of those states placed on the $(n,M^2)$
trajectories. In the region of 2000~MeV three $n\bar{n}$-dominant
resonances, $f_2(1920)$, $f_2(2000)$ and $f_2(2020)$, are seen,
while on the $(n,M^2)$-trajectories there are only two vacant
places. This means that one state is obviously superfluous from
the point of view of the $q\bar q$-systematics, i.e. it has to be
considered as exotics.

There exist various arguments in favour of the assumption that
$f_2(2000)$ is generated by a glueball. Still, it can not be a pure
gluonic $f_2(2000)$ state: it follows from the $1/N$ expansion rules
\cite{t'hooft,ven} that the gluonic state $(q\bar q)$ mixes with
quarkonium systems $(gg)$ without suppression. The problem of the
mixing of $(gg)$ and $(q\bar q)$ systems is discussed in Section 4,
where we present also the relations between decay constants of a
glueball into two pseudoscalar mesons $glueball\to PP$ and into two
vector mesons $glueball\to VV$.

In Section 5 we demonstrate that just $f_2(2000)$ is the glueball.

In \cite{PR-exotic} it was pointed out that an exotic state has to
be broad. Indeed, if an exotic resonance occurs among the standard
$q\bar q$-states, they overlap, and their mixing becomes possible
due to large distance transition: the $resonance\,(1)\to real\
mesons \to resonance\,(2)$. Owing to these transitions, an exotic
meson accumulates the widths of its neighbouring resonances. The
phenomenon of the accumulation of widths was studied in the scalar
sector near 1500~MeV \cite{APS-PL,AAS-PL}. In \cite{AAS-PL}, a model
of mixing of the gluonium $gg$ with the neighbouring quarkonium
states was considered. It was demonstrated that, as a result of
mixing, it is precisely the gluonium state which transforms into a
broad resonance. The reason is that $q\bar q$ states being
orthogonal to each other mix weakly, while the gluonium mixes with
neighbouring $q\bar q$ states without suppression. Therefore, the
gluonium "dives" \ more rapidly into the complex $M$-plane. The
mixing of states is always accompanied by a repulsion of the
corresponding poles: when poles are in the complex $M$-plane at
approximately the same Re $M$, this repulsion results in "sinking"
one of them into the region of large and negative Im $M$ and
"pulling" \  others to the real $M$-axis.

Hence, the large width of $f_2(2000)$ can indicate that this state
is an exotic one. Strictly speaking, this fact is not sufficient
to prove that $f_2(2000)$ is a glueball. At the moment a variety
of versions for exotic mesons is discussed; these are
$q\bar q g$ hybrids as well as multiquark states (see, e.g.
\cite{book,Bugg,ufn04} an references therein). Thus, in order to
fix $f_2(2000)$ as a glueball, it is of great importance to
investigate the decay couplings and prove that they satisfy
relations characterising the glueball. The coupling constants for
the transitions
$$ f_2(1920),f_2(2000), f_2(2020), f_2(2240),
f_2(2300)\to\pi\pi, \eta\eta,\eta\eta' $$ are separated in
 \cite{AMNS,AS} on the basis of a partial wave analysis
\cite{Ani} \ carried out earlier. The coupling constants obtained in
\cite{AMNS,AS} indicate that only the decays $f_2(2000)\to
\pi^0\pi^0,\eta\eta,\eta\eta'$ obey the relations corresponding to a
glueball, while the decay constants for other resonances do not
fulfil such conditions. Note that the glueball decay couplings are
close to those for the $SU(3)$-flavour singlet, but, because of the
flavour symmetry violation caused by the strange quark, do not
coincide with them exactly.

Let us remind that there are two more arguments in favour of
the glueball nature of $f_2(2000)$: \\
(i) the Pomeron trajectory, determined on the basis of data on
high-energy hadron decays (see, e.g., \cite{kaid,land,dakhno}),
indicates that a tensor glueball has to have a mass of the order
of $1.7-2.3\, GeV$; \\
(ii) lattice calculations \cite{lattice} lead to a similar value,
$M_{2^+-glueball}\sim 2.3-2.5\, GeV$.

We have one, sufficiently general, argument against the
interpretation of $f_2(2000)$ as an exotic $q\bar q g$ or $qq\bar
q\bar q$ state: the absence of any serious facts confirming their
existence. Indeed, if such states existed, we would see a large
number of them in the mass region above $1500$ MeV. Moreover, we
could observe not only exotic states; the number of resonances
with "normal"$\,$quantum numbers would also be seriously increased.
However, the systematics of quarks on the $(n,M^2)$-plane does not
reveal such an increase: almost all observed resonances can be
interpreted as $q\bar q$ states (see \cite{book}, Chapter 5).
Apparently, Nature does not like coloured multiparticle objects.
The same conclusion follows from the systematisation of baryons:
experimental data give a much smaller amount of excited states,
than calculations in the framework of a three-quark model
\cite{petry} do. One gets the impression that excited baryons are
rather quark--diquark systems (see discussions in
\cite{book,klempt-baryon}.

Owing to the $1/N$ expansion rules, the gluonium component is
relatively small in the quark state $f_2$: its probability is
suppressed as $1/N_c$. In Section 6 we determine the mixing angle
of the $n\bar n$ and $s\bar s$ components in the quark
$f_2$-mesons, making use of the relations between the decay
constants $f_2\to \pi\pi,\eta\eta,\eta\eta'$. Also, we estimate
the possible changes in the mixing angle as a consequence of a
gluonium component in the quark state $f_2$.

In the Conclusion, we discuss the situation in the glueball
sector.

Up to now, two glueball states, the scalar meson $f_0(1200-1600)$
an the tensor $f_2(2000)$ are observed. Both states are broad
ones, and the coupling constants corresponding to their decays
into pseudoscalar mesons (channels $glueball\to PP$) satisfy just
the relations characterising the glueball. The next states which
are of interest are radial excitations of the scalar and tensor
gluonia, and the pseudoscalar glueball. Taking the Pomeron
trajectories on the $(J,M^2)$-plane as a basis, we predict the
masses of excited scalar and tensor glueball states.

\section{Analysis of the data for tensor mesons }

We demonstrate here the results obtained from the data analysis
used in performing the systematisation of the $f_2$ resonances and
extracting the decay couplings $f_2\to \pi\pi,\eta\eta,\eta\eta'$.

\subsection{ L3 data on the $\gamma\gamma\to \pi^+\pi^-\pi^0$ reaction}

In this reaction the $\gamma\gamma$ channel couples only to states
with C=+1 parity; $3\pi$ has a negative G-parity. For a $q\bar q$
system one has $G=Ce^{i\pi I}$, so the $I=1$ quark-antiquark
states are produced only in the $\gamma\gamma$ channel. Due to
$C$-parity conservation in neutral decay modes, only $f$-states
with ($J^{PC}=0^{++}$, $2^{++}$, $4^{++}...$) are produced in the
$\pi^+\pi^-$ channel. In the $\pi^{\pm}\pi^0$ channel only isovector
mesons with $J^{PC}=1^{--}$, $3^{--}\ldots\,$ are produced.

\begin{figure}[t!]
\centerline{\epsfig{file=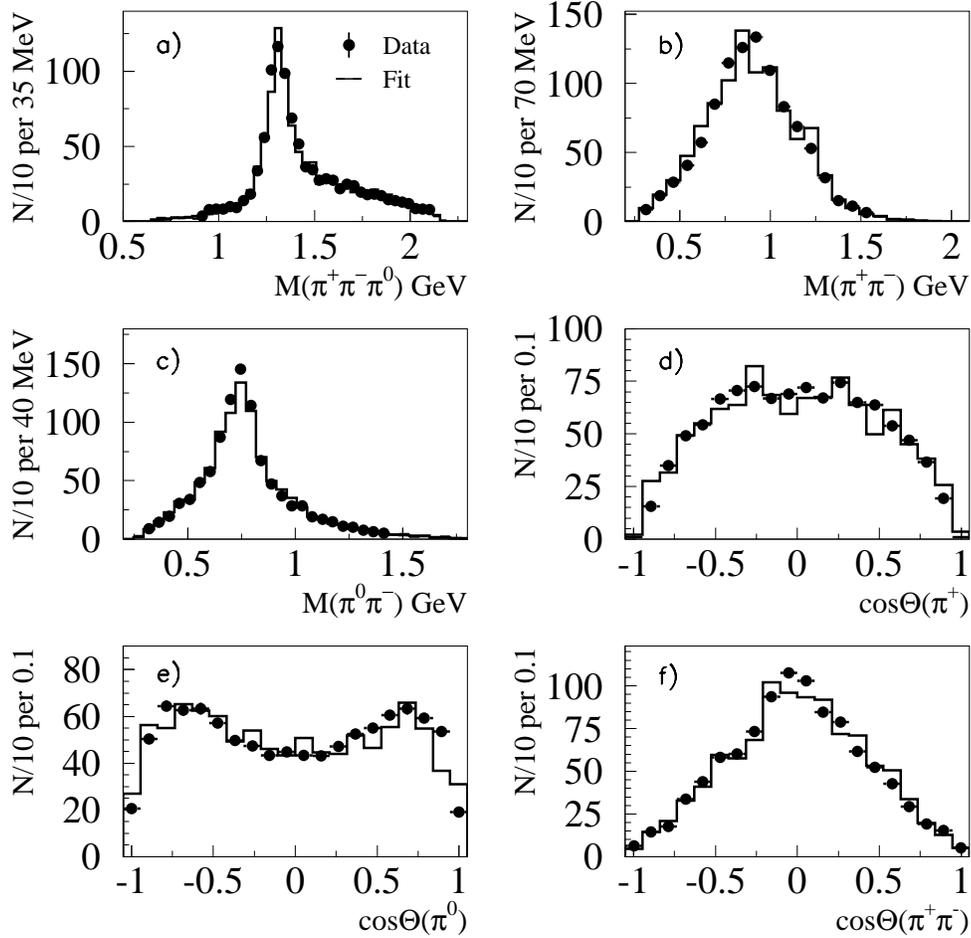,width=15cm}} \vspace{-1.cm}
\caption{ Reaction $\gamma\gamma\to \pi^+\pi^-\pi^0$. a)
$\gamma\gamma$ mass spectrum, b,c) $\pi^+\pi^-$ and $\pi^{\pm}\pi^0$
mass spectra, d,e) the angular distributions of charged and
neutral pions in the c.m.s. of the reaction, f) the angular
distribution between charged and neutral pion in the c.m.s. of two
charged pions.}
\label{m_proj}
\end{figure}

\begin{figure}[t!]
\centerline{\epsfig{file=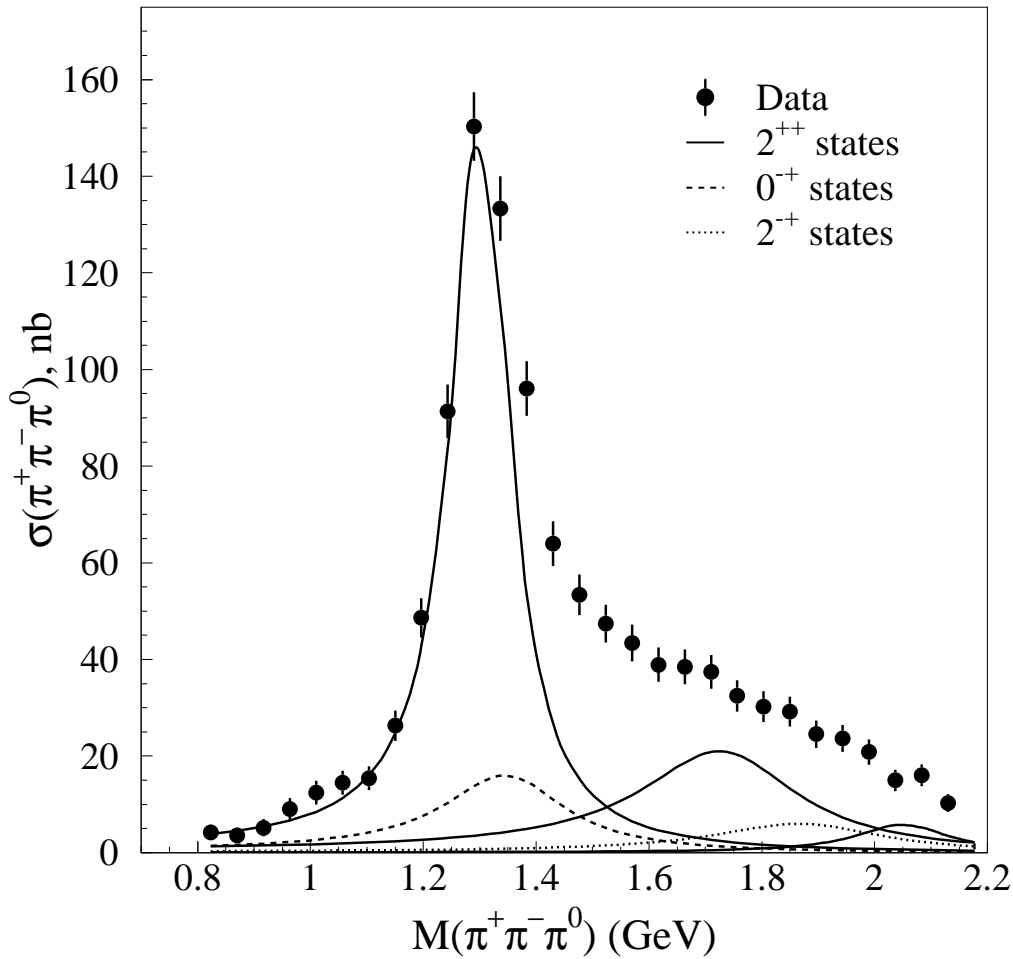,width=15cm}} \vspace{-1.cm}
\caption{ Reaction $\gamma\gamma\to \pi^+\pi^-\pi^0$. The
contribution of different resonances listed in Table 1 to the
cross section: full curves - $2^{++}$ states, dashed curves -
$0^{-+}$ states and dotted curves - the contribution of $2^{-+}$
states.}
\label{gamma_c}
\end{figure}

The $\gamma\gamma$ mass distribution is dominated by the production
of the $a_2(1320)$ resonance, see Fig. \ref{m_proj}a.
One can see a prominent structure in the mass region 1.6-1.8 GeV
as well as a possible contribution at the $a_2(1320)$ signal.

The $\pi^+\pi^-$ mass distribution is shown in Fig.~\ref{m_proj}b.
There are no clear signals in the data coming from well known
narrow scalar-isoscalar states $f_0(980)$ and $f_0(1500)$.
Indeed, the partial wave analysis shows very small contributions
of these mesons; such decay modes were omitted in the final fit.

A signal coming from $f_2(1275)\pi^0$ is observed at high
$\gamma\gamma$ energies; this is important to describe the
two-pion mass spectrum and angular distributions.

The $\pi\pi\to \pi\pi$ S-wave amplitude has a broad component
which covers the mass region from the $\pi\pi$ threshold up to 2
GeV. Such a component is introduced in the present analysis and is
parametrized in two different ways.

\begin{table}[b!]
\begin{center}
\caption{Masses, total widths and the partial width
$\Gamma_{\gamma\gamma\to \pi\pi\pi}$ for the resonances observed in
the reaction $\gamma\gamma\to \pi^+\pi^-\pi^0$. }
~\\
\begin{tabular}{|l|l|l|l|}
\hline Resonance & M (MeV) & $\Gamma$(MeV) &
$\Gamma_{\gamma\gamma}$Br$(3\pi)$(keV) \\
\hline
$a_2(1320)$ & $1302\pm 3\pm 6$ & $118\pm 6\pm 10 $ & $ 0.65\pm0.05$\\
$a_2(1730)$ & $1725\pm 25\pm 10$ & $340\pm 40$ & $0.34^{+0.15}_{-0.06}$ \\
$\pi(1300)$ [1] & $1350\pm 40$ & $320\pm 50$ & $ \le 0.8$ \\
$2^{-+}$ & $1870\pm 60$ & $325\pm 40$ & $ 0.15\pm 0.03$ \\
$\pi_2(1670)[1]$ & $1670$ & $260$ & $\le 0.1$ \\
\hline
\end{tabular}
\end{center}
\label{3pi}
\end{table}

The first parametrisation is taken from \cite{bszz}. It was
introduced to describe the CERN-Munich data on ($\pi^-p\to
\pi^+\pi^- n$) \cite{cern} and the Crystal Barrel data on
proton-antiproton annihilation into $3\pi^0$ and $2\eta\pi^0$
channels simultaneously. To simulate a possible s-dependence of the
vertex (which can be important for this very broad state), we use
the method suggested in \cite{bszz}.

The second parametrisation was used in \cite{kmat}. It covers the
mass region up to 1.9 GeV and describes, in the framework of the
P-vector/K-matrix approach, a much larger number of two- and
three-body reactions. To avoid an over-parametrisation of the fit,
we vary only the production couplings of the two lowest K-matrix
poles.

The main signal in the $\pi^{\pm}\pi^0$ mass spectrum is due to the
production of $\rho(770)$. There is very little structure in the
region higher than 1 GeV (see Fig.~\ref{m_proj}c) and the signal
is almost zero at masses above 1.5 GeV. Because of this, neither
$\rho_3(1690)$, nor $\rho(1770)$ has to be introduced in the
analysis. A contribution from $\rho(1450)$ is found to be useful
to describe the data: however, this state is quite broad and
possibly simulates a non-resonant two pion production in this
channel.

In the $\pi^+\pi^-\pi^0$ spectrum one can see a strong signal
coming from $a_2(1320)$; the characteristics of this resonance
were defined with high precision by the VES collaboration
\cite{VES}. It is not surprising that the $\gamma\gamma\to3\pi$
data are dominated by the production of the $a_2(1320)$ state,
since this resonance has the highest spin in the mass region below
1.6 GeV (the $\gamma\gamma$ cross section is proportional to
$(2J+1)$) and it is a ground $q\bar q$ state with the radial
quantum number $n=1$. Indeed, the $\gamma\gamma\to resonance$
production amplitude is a convolution of the photon and the
quark-antiquark resonance wave functions
\cite{AMN,anis_nik,gamma}. This provides a suppression of nearly
an order for the production of the radially excited states
$(n\geq 2)$ \cite{anis_nik}. Nevertheless, there is a manifest
contribution of the higher tensor state. While $a_2(1320)$ decays
practically only into the $\rho(770)\pi$ channel, the second
tensor state decays almost equally to
$\rho(770)\pi$ and $f_2(1275)\pi$. This fact allows us to identify
this state with a good accuracy.

The solution reveals quite a large contribution coming from the
$0^{-+}$ partial wave decaying into $f_0\pi$. There is, however, a
problem in distinguishing it from the experimental background:
such a decay, giving S-wave amplitudes in all decay channels,
provides very smooth structures in mass distributions and,
moreover, these amplitudes do not interfere with the tensor
amplitude.

The contribution of the $2^{-+}$ state is found to be quite small
when fitted to the $\pi_2(1670)$ state (see Table 1). If it is
fitted with free Breit-Wigner parameters, it is always optimised
at higher masses ($\sim 1870$ MeV).

The $ \pi^+\pi^-\pi^0$ spectrum and the contributions of
resonances with different $J^{PC}$ in the final solution are shown
in Fig.~\ref{gamma_c}. Masses, total widths, the
$\Gamma_{\gamma\gamma}$ partial width and the branching ratio into
$3\pi$ are listed in Table 1 for the considered resonances.

\subsection{ L3 data on the reaction $\gamma\gamma\to K_S K_S$ \cite{L3}}

Important data for the identification of the tensor mesons were
obtained by the L3 collaboration on the reaction $\gamma\gamma\to
K_SK_S$. Only states with even spin $J$ and positive parities
$P\!=\!C\!=\!+$ contribute to two neutral pseudoscalar particles,
what reduces the possible partial waves drastically. The $(2J+1)$
factor in the cross section favours the dominant production of the
tensor states. The $4^{++}$ states are produced at high energies
$(M\ge 1.9\,GeV)$.

The meson states consisting of light quarks $u$, $d$ $s$ form meson
nonets: three isospin 1, four isospin $1/2$
and two isoscalar states. The isoscalar states can be a mixture of
$n\bar n=(u\bar u+d\bar d)/\sqrt 2$ and $s\bar s$ components. The
decay of $(I=0)$ and $(I=1, I_3=0)$-states into two kaons is
defined by the production of a new $s\bar s$ pair (an $s$-quark
exchange process) and has the following structure for different
isospins: $(u\bar u+d\bar d)/\sqrt 2\to K^+K^- + K^0 \bar K^0$ for
I=0, and $(u\bar u- d\bar d)/\sqrt 2\to K^+K^- - K^0 \bar K^0$ for
I=1. As a result, a strong destructive interference occurs
between the $f_2(1275)$ and $a_2(1320)$ mesons which is very
sensitive to the mixing angle of the isoscalar state. The flavour
content of isoscalar-scalar resonances belonging to the same nonet
can be written in the form
\be
f_2(q\bar q) = & n\bar n \cos \varphi & + s\bar s~\sin \varphi
\;,\\
f'_2(q\bar q) = & -n\bar n \sin \varphi & + s\bar s~\cos \varphi
\;.
\ee

Although the production of states with dominant $s\bar s$
components is suppressed by the smaller $\gamma \gamma$ coupling,
these states usually have a larger branching ratio for the decay
into the $K_S K_S$ channel, and can contribute appreciably into
the total cross section.

\begin{figure}
\centerline{\epsfig{file=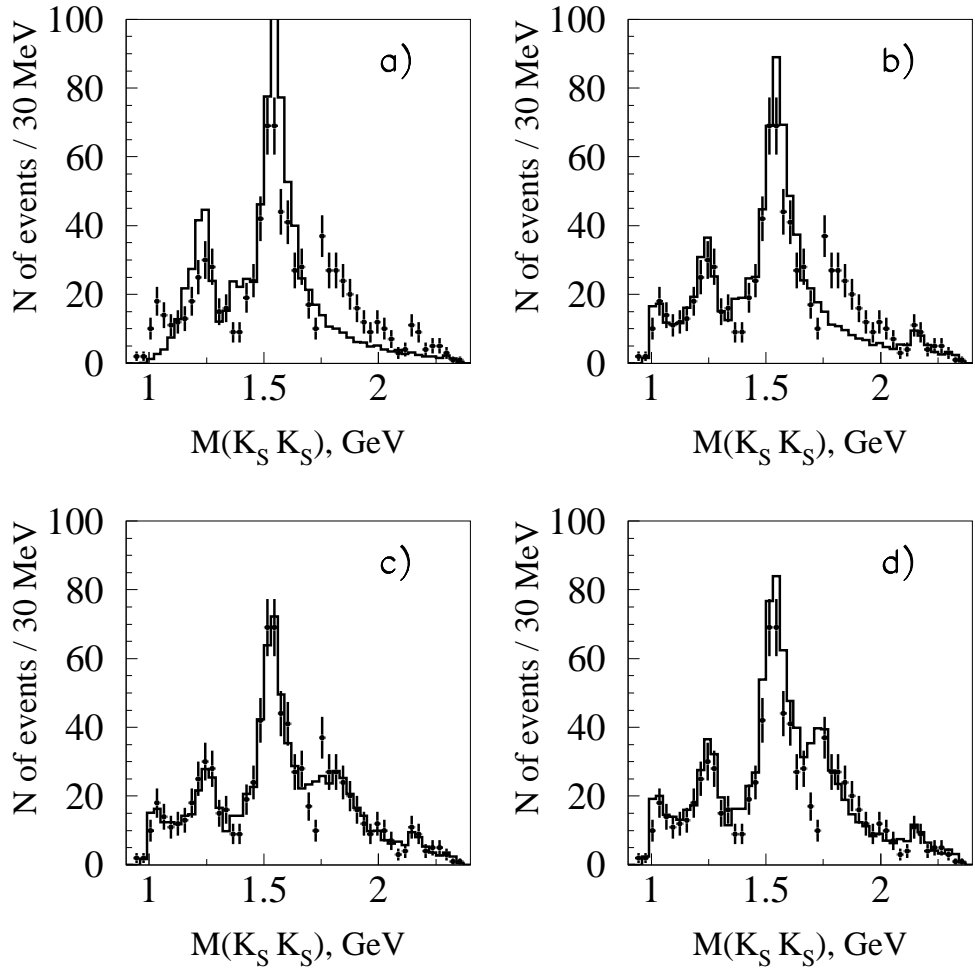,width=15cm}}
\caption{Reaction $\gamma\gamma\to K_S K_S$: a) the description of
the data with the first tensor nonet, b) same as (a) plus scalar
states and a $4^{++}$ state, c) same as (b) but the scalar states
with free parameters , d) same as (b) but $f_0(1710)$ state with
parameters fixed from the latest BES results.} \label{ksks_all}
\end{figure}

There is no doubt about the nature of tensor resonances below 1600
MeV. The partial wave analysis showed that tensor resonances are
produced dominantly in the $^5S_2$ $\gamma\gamma$ state, which was
predicted by model calculations \cite{anis_nik}. The model
gives directly the ratio between $^5S_2$ and $^1D_2$ waves and
the $\gamma\gamma$ couplings. These values can be introduced in
the analysis without changing the quality of the description.
The data with the lowest tensor states are shown in
Fig.~\ref{ksks_all}a.

We have found that the $0^{++}$ partial wave can be fitted well in
the framework of the P-vector/K-matrix approach. The form of the
$0^{++}$ contribution in $\gamma\gamma\to K\bar K$ follows closely
the $\pi\pi\to K\bar K$ cross section, which is not surprising:
the production coupling to the $s\bar s$ system is strongly
suppressed in both reactions. The values of the coupling constants
of the reactions $\gamma\gamma\to f_0,a_0$ agree well with those
given in the calculations \cite{anis_nik}: in the same way as in
the tensor sector, the calculated $\gamma\gamma$ couplings can be
used directly, not damaging the quality of the description.

A contribution of $4^{++}$ states is observed in high energy
angular distributions. Some broad and some rather narrow
components are seen in this wave. The broad state can be
associated with $n\bar n$ and the narrow one with a $4^{++}$
$s\bar s$ state. The description of the data with tensor, scalar
and $4^{++}$ states is shown in Fig.~\ref{ksks_all}b.

There is a clear resonance structure in the $1750$ MeV region. The
angular distribution in this region follows very closely the
$(1-\cos^2\Theta)^2$ shape which corresponds to the
$^5S_2\gamma\gamma$ production of the tensor meson. However, the
acceptance decreases rapidly in the forward and backward
directions providing a very similar dependence. Due to such a
behaviour and to the not too high statistics, the partial wave
analysis produces almost the same angular distribution for a
tensor state and for a scalar state. Still, the fit using a scalar
state fails to reproduce the structure in the 1700-1800 MeV mass
region. The description of the data with the best $\chi^2$ is
shown in Fig.~\ref{ksks_all}c. The mass was optimised to $1805\pm
30$ MeV and the width to $260\pm 30$ MeV. With such a mass and
width, the $f_0$ state can describe the slope in mass distribution
above 1800 MeV. If the mass and width of the scalar state are
fixed at the BES result \cite{bes} $M=1740$ MeV, we obtain a
description shown in Fig.~\ref{ksks_all}d.

The main problem in a fit with a $f_0$ state is that there is no way
to reproduce the dip in the 1700 MeV region and the slope above 1800
MeV using any (even very sophisticated) parametrisation of the
resonance width. A $f_0$ state can interfere only with the $^1D_2$
component of a $2^{++}$ state, and this partial wave is very small
in the data. Consequently, $f_0$ and $f_2$ contributions practically
do not interfere and it is impossible to create a dip in the data.
Contrary to this, a tensor resonance interfering with the tails of
other tensor states naturally produces a dip and a good description
of this mass region.

All isoscalar and isotensor states can contribute to the
$\gamma\gamma\to K_SK_S$ cross section; this situation offers a
very good possibility to study the reaction on the basis of the
nonet classification. With $SU(3)$ relations imposed, the only
parameters to fit the data for the first three states are masses,
widths, the mixing angle and $SU(3)$ violation factors. We found
all masses and widths for the members of the first tensor nonet to
be in a very good agreement with PDG.

\begin{figure}
\centerline{\epsfig{file=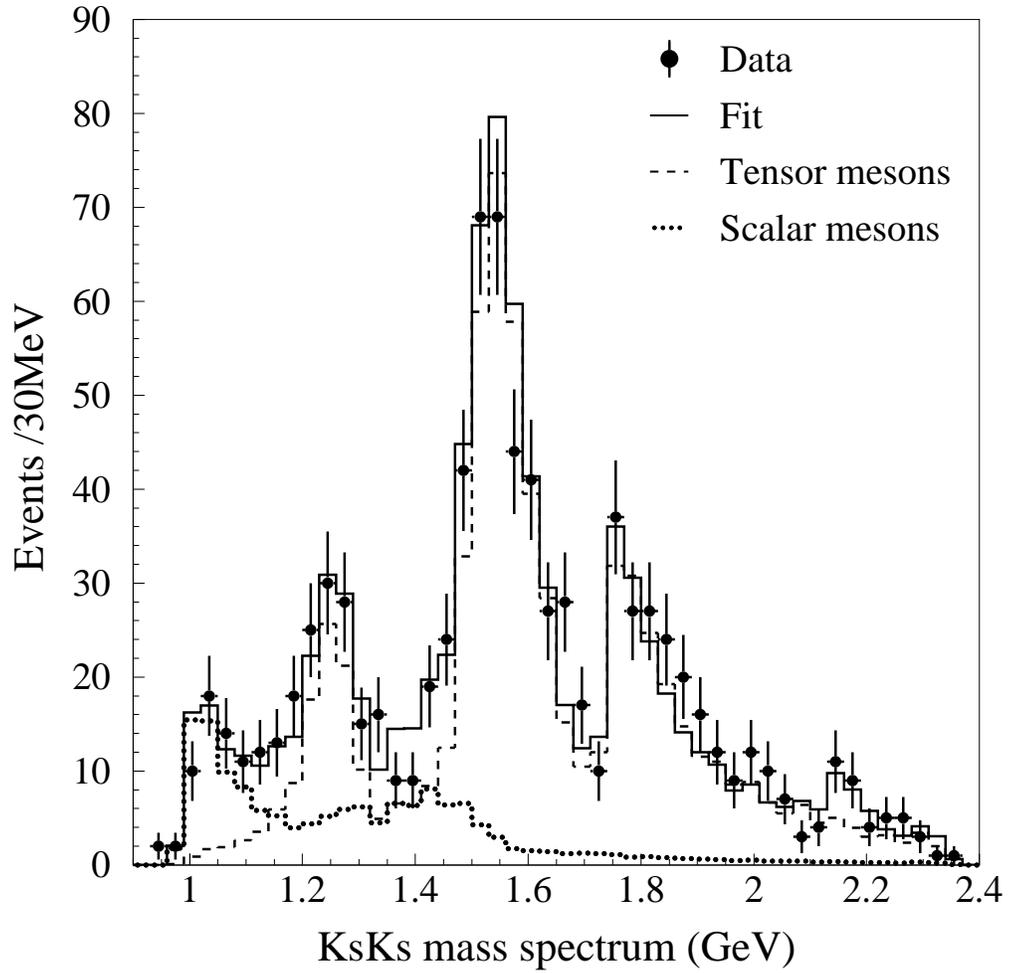,width=15cm}}
\caption{The description of the reaction $\gamma\gamma\to K_S K_S$
(full curve), and different contributions to the cross section:
dashed curve -- from the tensor states and dotted curve -- from
the scalar states} \label{ksks}
\end{figure}

To describe the second nonet, we fixed parameters for $f_2(1560)$
from the Crystal Barrel results and for $a_2(1700)$ from the
latest L3 analysis of the $\pi^+\pi^-\pi^0$ channel \cite{L3-3pi}.
At a given
mixing angle the nonet coupling was calculated to reproduce the
$\pi\pi$ width of 20-25 MeV for the $f_2(1560)$ state.
We found a very good description of the data with $SU(3)$
relations imposed, similar to the fit described in the previous
section. The masses, widths, radii, $K\bar K$ couplings, mixing
angles and partial widths of the states are given in
Table 2;
the description of the data is shown in Fig.~\ref{ksks}.

The $f_2(1750)$ state has a mass $1755\pm 10$ MeV and a total
width $67\pm 12$ MeV. Its $K\bar K$ width is $23\pm 7$ MeV; the
rest of the width is likely to be defined by the $K^*\bar K$
channel. This resonance destructively interferes with the tail of
the $f_2'(1525)$ state, creating a dip in the mass region 1700
MeV. When the sign of the real part of the $f_2(1750)$ amplitude
changes, this interference becomes positive, producing a clear
peak in the data.

A serious problem appears in the description of the data in the
framework of the nonet approach, if the peak at 1750 MeV is
assumed to be owing to a scalar state. If this state is a nonet
partner of one of the known states, e.g. $f_0(1370)$ or
$f_0(1500)$, then the calculated signal is too weak to fit the
data. If the $K\bar K$ coupling of this scalar state is fitted
freely, we find that it must be about four times larger than the
total width of the resonance. This is due to the $2J+1$
suppression factor and to the absence of a positive interference
with the tail of $f'_2(1525)$ which boosts the peak in the case of
a tensor state. These are problems additional to those connected
with the description of the dip in the 1700 MeV region.

\begin{table}
\begin{center}
\caption{Parameters of resonances observed in the reaction
$\gamma\gamma\to K_SK_S$. The values marked with stars are fixed
by using other data.}
~\\
\begin{tabular}{|c|c|c|c|c|c|c|}
  \hline
  ~ & \multicolumn{3}{c|}{First nonet} &
  \multicolumn{3}{c|}{Second nonet} \\
  \hline
  ~ & $a_2(1320)$ & $f_2(1270)$ & $f'_2(1525)$
         & $a_2(1730)$ & $f_2(1560)$ & $f_2(1750)$ \\
  \hline
 Mass (MeV) & $1304\pm 10$ & $1277\pm 6$ & $1523\pm 5$
            & $1730^*$ & $1570^*$ & $1755\pm 10$ \\
\hline Width (MeV) & $120\pm 15$ & $195\pm 15$ & $104\pm 10$
            & $340^*$ & $160^*$ & $67\pm 12$ \\
\hline
 $g^L$ (GeV) & $0.8\pm 0.1$ & $0.9\pm 0.1$ & $1.05\pm0.1$
              & \multicolumn{3}{c|}{$0.38\pm 0.05$} \\
\hline
$\varphi$ (deg) & \multicolumn{3}{c|}{$-1\pm 3$}
              & \multicolumn{3}{c|}{$-10^{+5}_{-10}$} \\
\hline
\end{tabular}
\end{center}
\label{tab:ksks}
\end{table}

\begin{figure}
\psfig{file=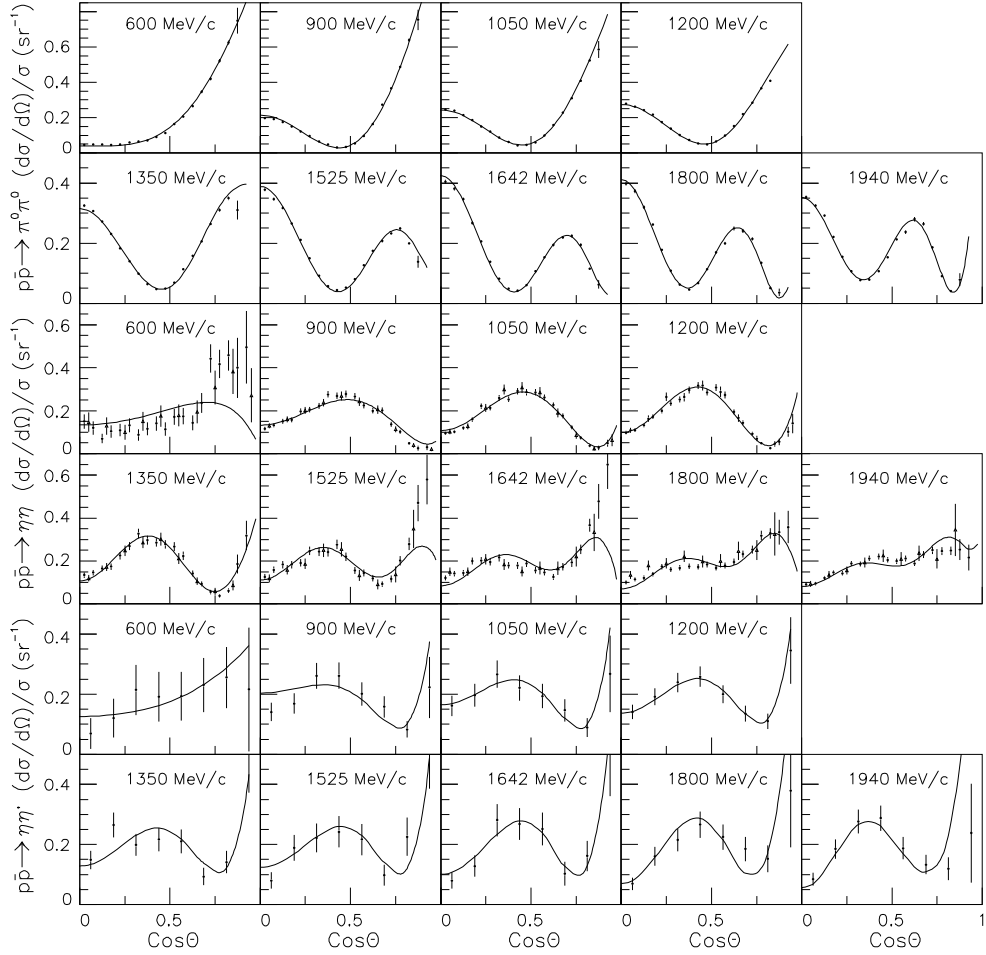,width=15cm}
\caption{ Angular distributions in the reactions $p\bar p\to
\pi\pi,\eta\eta,\eta\eta'$ and their fit within resonances of
Eq.(2).}
\end{figure}

\begin{figure}
\psfig{file=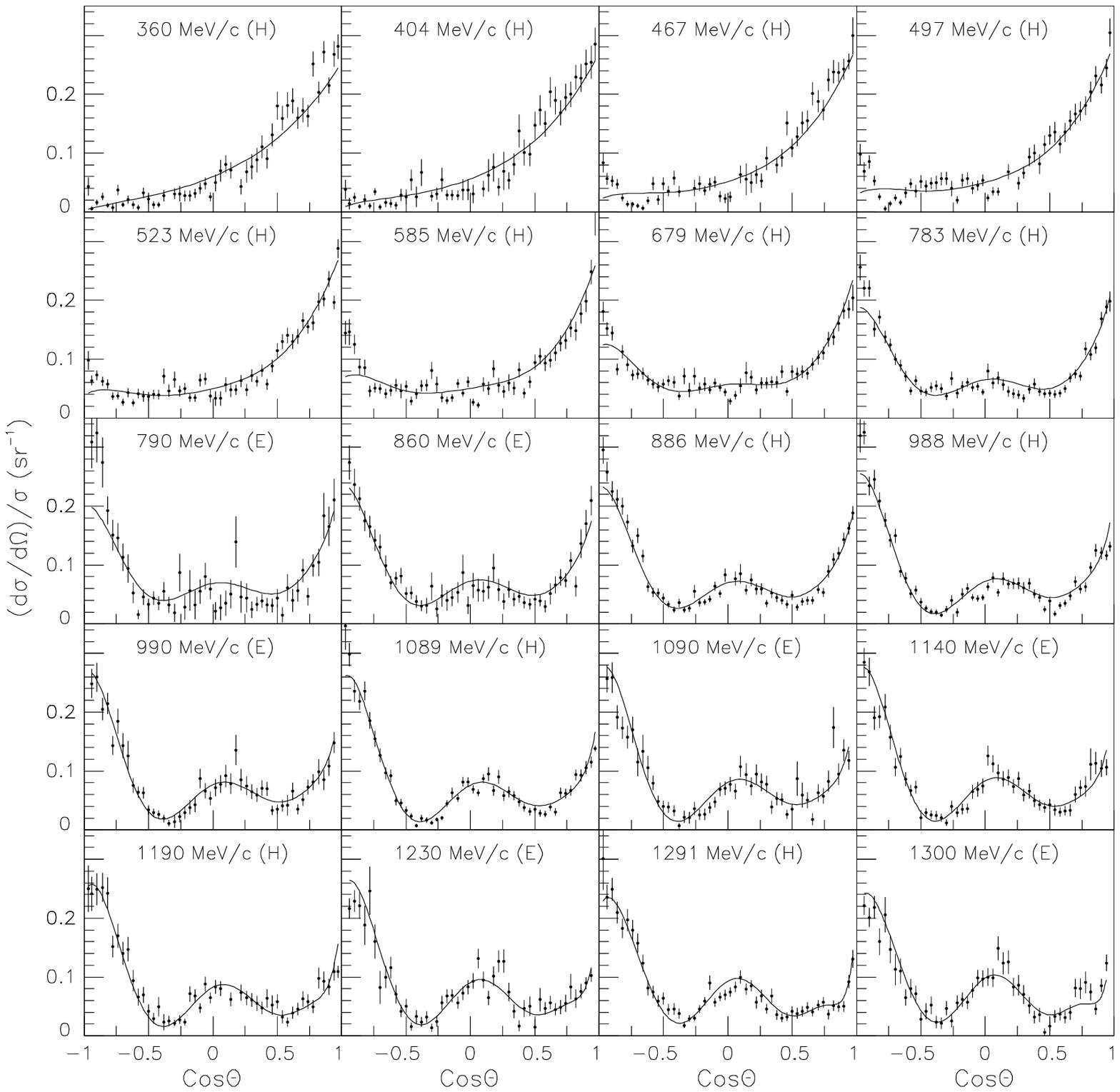,width=15cm}
\caption{ Differential cross sections in the reaction $p\bar
p\to\pi^+\pi^-$ [42] at proton momenta 360-1300 MeV and their fit
within resonances of Eq. (2).}
\end{figure}

\begin{figure}
\psfig{file=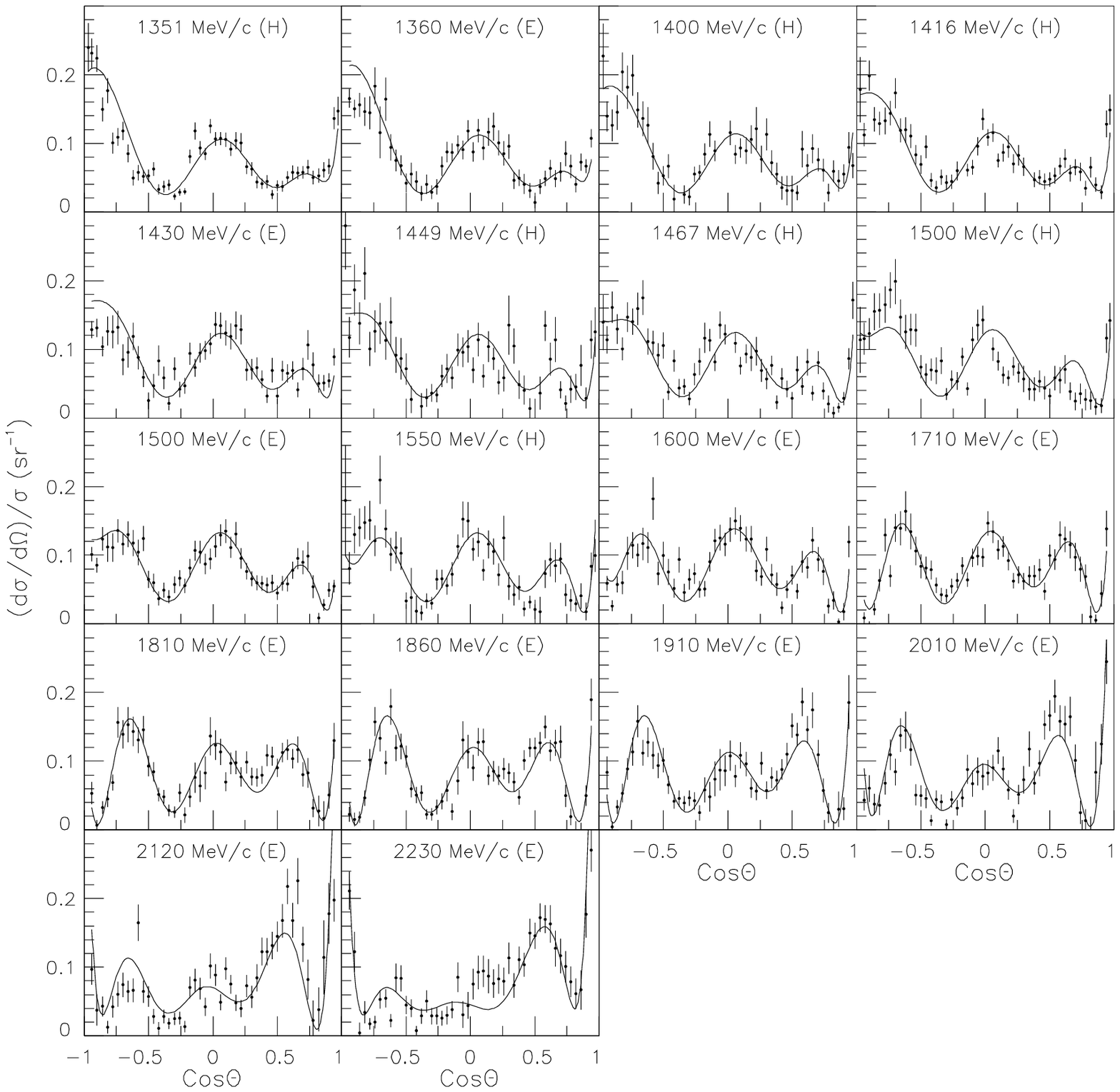,width=15cm}
\caption{ Differential cross sections in the reaction $p\bar
p\to\pi^+\pi^-$ [42] at proton momenta 1350-2230 MeV and their fit
within resonances of Eq.(2). }
\end{figure}

\begin{figure}
\psfig{file=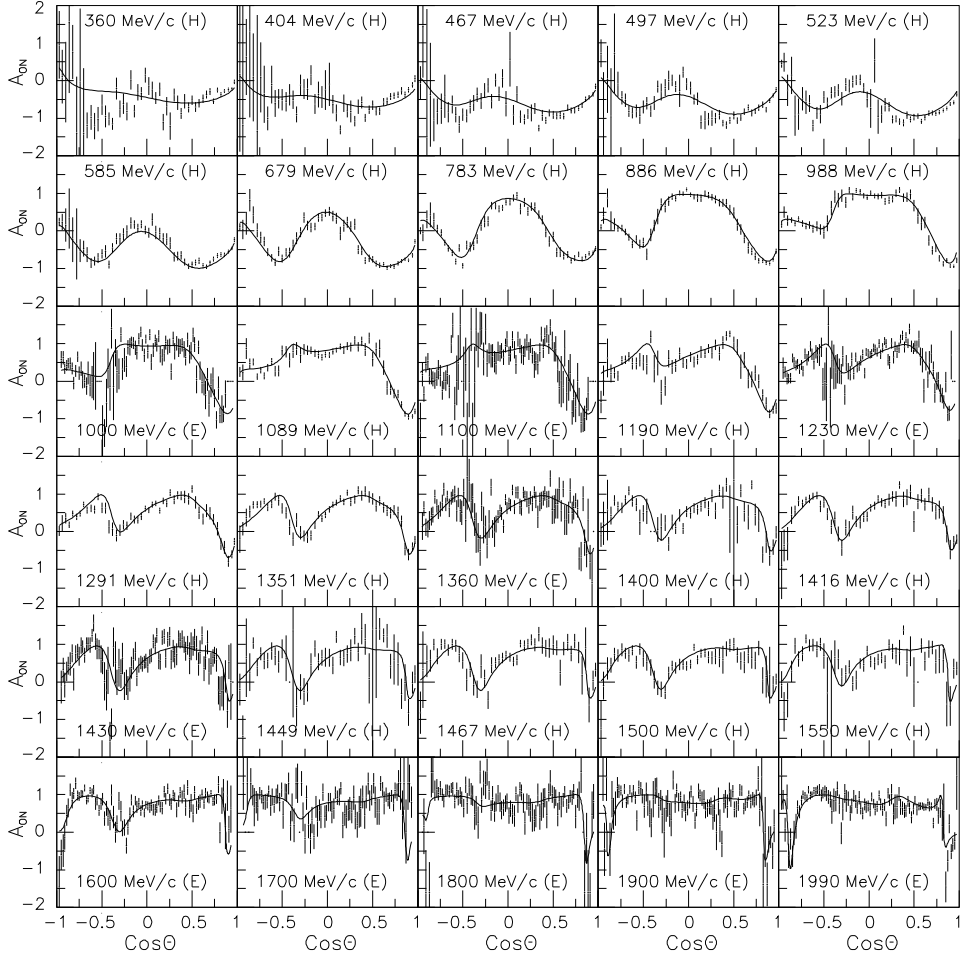,width=15cm}
\caption{ Polarisation in $p\bar p\to\pi^+\pi^-$ [42] and its fit
within resonances of Eq. (2). }
\end{figure}

\begin{figure}
\centerline{\epsfig{file=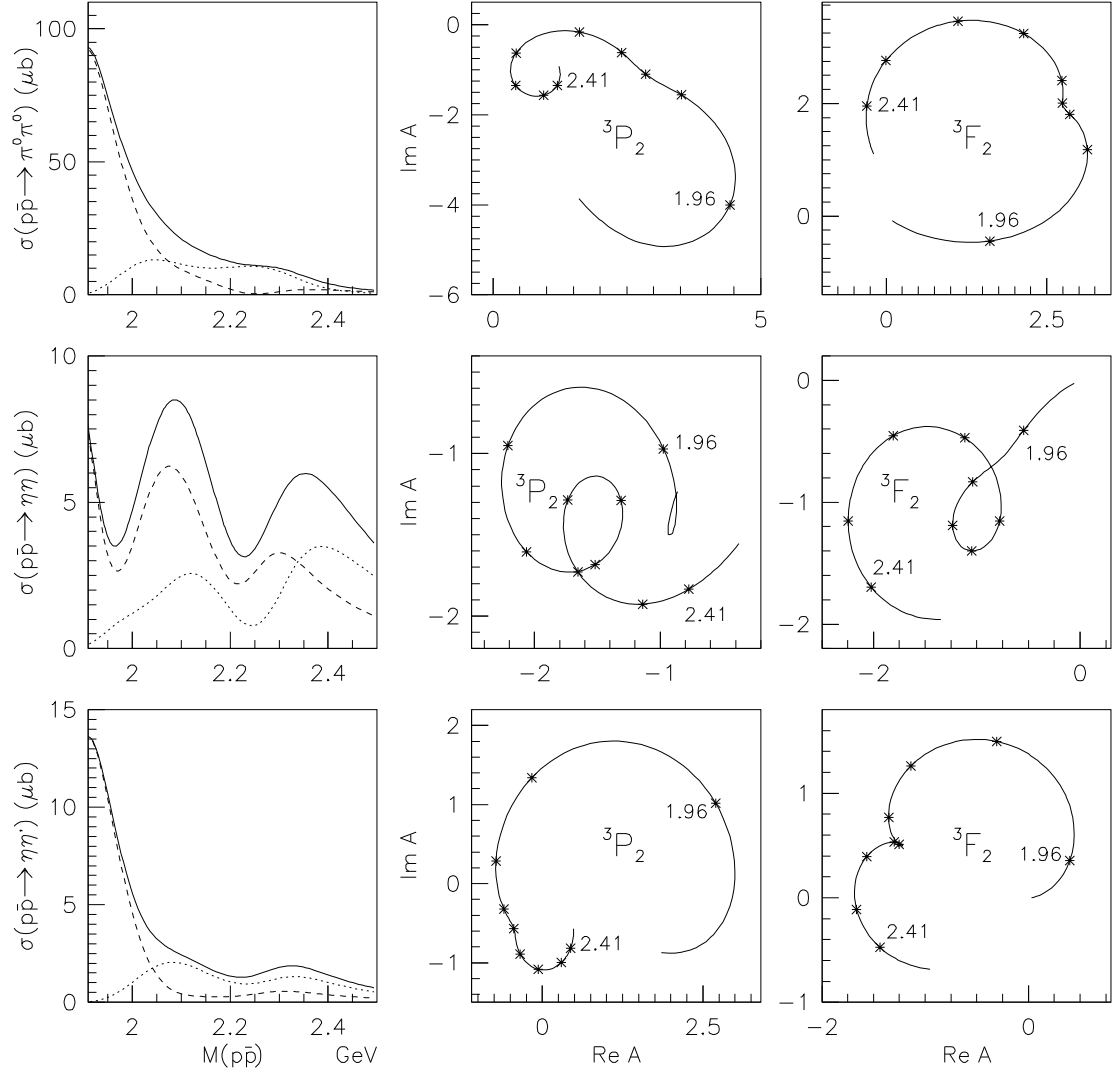,width=17cm}}
\caption{Cross sections and Argand-plots for $^3P_2$ and $^3F_2$ waves
in the reaction $p\bar p\to\pi^0\pi^0,\eta\eta,\eta\eta'$. The
upper row refers to $p\bar p\to\pi^0\pi^0$: we demonstrate the
cross sections for $^3P_2$ and $^3F_2$ waves (dashed and dotted
lines, respectively) and the total $(J=2)$ cross section (solid
line) as well as Argand-plots for the $^3P_2$ and $^3F_2$ wave
amplitudes at invariant masses $M=1.962$, $2.050$, $2.100$,
$2.150$, $2.200$, $2.260$, $2.304$, $2.360$, $2.410$ GeV. The
figures in the second and third rows refer to the reactions $p\bar
p\to\eta\eta$ and $p\bar p\to\eta\eta'$.}
\end{figure}

\subsection{Data for proton-antiproton annihilation in flight
$p\bar p\to \pi\pi,\eta\eta,\eta\eta'$}

The $p\bar p$ annihilation in flight gives information about
resonances with $M>1900$ MeV. High statistical data taken at
antiproton momenta 600, 900, 1150, 1200, 1350, 1525, 1640, 1800
and 1940 MeV/c were used for the analysis
$p\bar p\to \pi^0\pi^0$, $\eta\eta,$ $\eta\eta'$ \cite{Ani}.
The combined analysis
was performed together with $\bar p p\to \pi^+\pi^-$ data obtained
with a polarised target \cite{Eisen}. Five tensor states are
required to describe the data, $f_2(1920)$, $f_2(2000)$, $f_2(2020)$,
$f_2(2240)$, $f_2(2300) $:
\be \label{2.3}
{\rm Resonance} & {\rm Mass (MeV)} & {\rm Width (MeV)} \\
f_2(1920) & 1920\pm 30 & 230\pm 40 \nn \\
f_2(2000) & 2010\pm 30 & 495\pm 35 \nn \\
f_2(2020) & 2020\pm 30 & 275\pm 35 \nn \\
f_2(2240) & 2240\pm 40 & 245\pm 45 \nn \\
f_2(2300) & 2300\pm 35 & 290\pm 50\, .
\ee

The description of the data is illustrated by Figs. 5,6,7 and 8.
In Fig. 9 we show the cross sections for $p\bar p\to
\pi^0\pi^0,\eta\eta,\eta\eta'$ in $^3P_2\bar p p$ and $^3F_2\bar p
p$ waves (dashed and dotted lines) and the total $(J=2)$ cross
section (solid line) as well as Argand-plots for the $^3P_2$ and
$^3F_2$ wave amplitudes at invariant masses $M=1.962$, $2.050$,
$2.100$, $2.150$, $2.200$, $2.260$, $2.304$, $2.360$, $2.410$ GeV.

The $\bar p p \to \pi^0\pi^0$, $\eta\eta$, $\eta\eta'$ amplitudes
provide the following ratios for the $f_2$ resonance couplings $
g_{\pi^0\pi^0}:g_{\eta\eta}:g_{\eta\eta'} $:
\be
\label{2.4}
& f_2(1920) & 1:0.56\pm 0.08:0.41\pm 0.07     \\
& f_2(2000) & 1:0.82\pm 0.09:0.37\pm 0.22  \nn \\
& f_2(2020) & 1:0.70\pm 0.08:0.54\pm 0.18  \nn \\
& f_2(2240) & 1:0.66\pm 0.09:0.40\pm 0.14  \nn \\
& f_2(2300) & 1:0.59\pm 0.09:0.56\pm 0.17.
\ee
These coupling ratios allow one to estimate the quarkonium-gluonium c
content of the $f_2$ - resonances.

\section{Systematisation of tensor mesons on the $(n,M^2)$
trajectories}

In \cite{syst} (see also \cite{book,ufn04}), the known $q\bar
q$-mesons consisting of light quarks ($q=u,d,s$) were put on the
$(n,M^2)$ trajectories, where $n$ is the radial quantum number of
the $q\bar q$ system with mass $M$. The trajectories for mesons
with various quantum numbers turn out to be linear with a good
accuracy:
\begin{equation}
\label{1} M^2\ =\ M^2_0+(n-1)\mu^2
\end{equation}
where $\mu^2=1.2\pm0.1\,\rm GeV^2$ is a universal slope, and $M_0$
is the mass of the lowest state with $n=1$.

In Fig. 10a we demonstrate the present status of the $(n,M^2)$
trajectories for the $f_2$ mesons (i.e. we use the results given
by \cite{Ani,LL,L3}). To avoid confusion, we list here the
experimentally observable masses. First, it concerns the
resonances seen in the $\phi\phi $ spectrum \cite{Etk}. In
\cite{LL} the $\phi\phi$ spectra were re-analysed, taking into
account the existence of the broad $f_2(2000)$ resonance. As a
result, the masses of three relatively narrow resonances are
shifted compared to those given in the compilation PDG \cite{PDG}:
$$
 f_2(2010)|_{PDG}\to f_2(2120)\ \cite{LL} , \quad
 f_2(2300)|_{PDG}\to f_2(2340)\ \cite{LL} , \quad
 f_2(2340)|_{PDG}\to f_2(2410)\ \cite{LL} .
$$
The trajectory for the $a_2$-mesons, Fig. 10b, is drawn on the
basis of the recent data \cite{AVAa2}.

\begin{figure}
\centerline{\epsfig{file=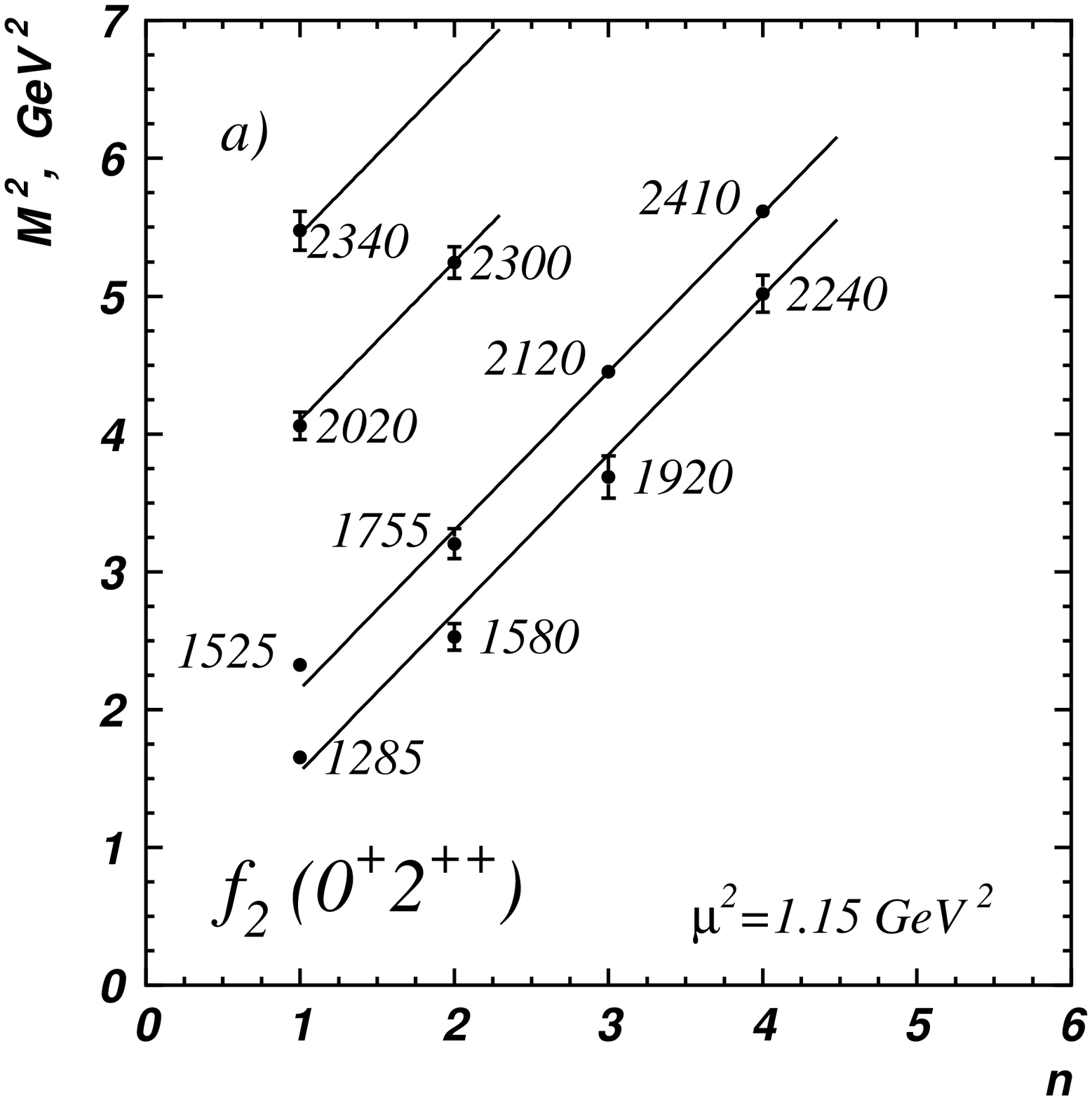,width=9cm}
\hspace{0.1cm}\psfig{file=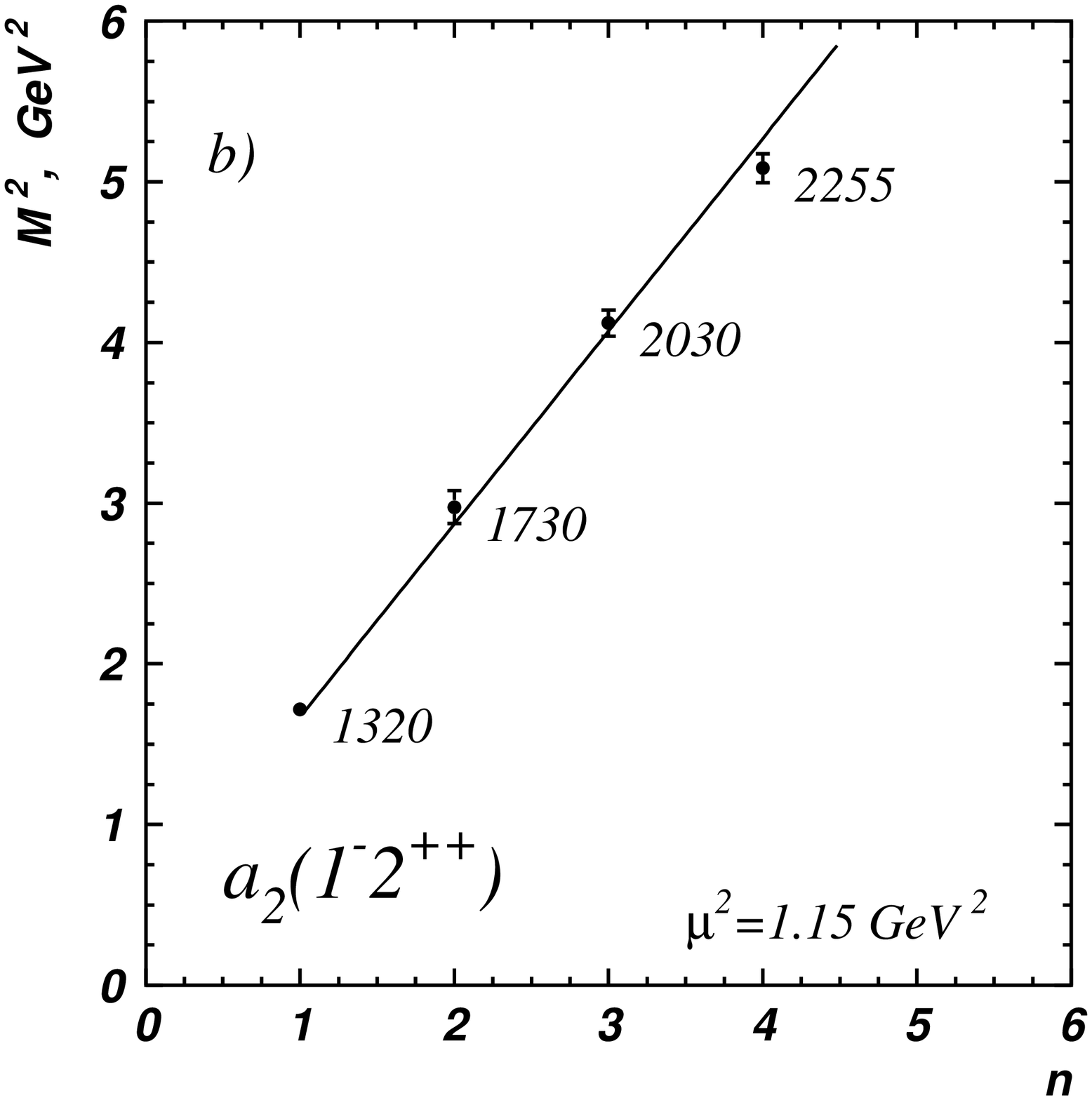,width=9cm}}
\caption{The $f_2$ and $a_2$ trajectories on the $(n,M^2)$ plane;
$n$ is the radial quantum number of the $q\bar q$ state. The
numbers stand for the experimentally observed $f_2$ and $a_2$
masses ($M$).}
\end{figure}

The quark states with ($I=0$, $J^{PC}=2^{++}$) are determined by two
flavour components $n\bar n$ and $s\bar s$ for which two states
$^{2S+1}L_J=\,^3P_2,\,^3F_2$ are possible. Consequently, we have
four trajectories on the $(n,M^2)$ plane. Generally speaking, the
$f_2$-states are mixtures of both the flavour components and the
$L=1,3$ waves. The real situation is, however, such that the lowest
trajectory [$f_2(1275)$, $f_2(1580)$, $f_2(1920)$, $f_2(2240)$]
consists of mesons with dominant $^3P_2n\bar n$ components (we
denote $n\bar n=(u\bar u+d\bar d)/\sqrt 2$), while the trajectory
$[f_2(1525)$, $f_2(1755)$, $f_2(2120)$, $f_2(2410)]$ contains mesons
with predominantly $^3P_2 s\bar s$ components, and the
$F$-trajectories are represented by three resonances
[$f_2(2020)$,$f_2(2300)$] and [$f_2(2340)$] with the corresponding
dominant $^3F_2n\bar{n}$ and $^3F_2s\bar s$ states. Following
\cite{glueball2}, we can state that the broad resonance $f_2(2000)$
is not part of those states placed on the $(n,M^2)$ trajectories. In
the region of 2000~MeV three $n\bar{n}$-dominant resonances,
$f_2(1920)$, $f_2(2000)$ and $f_2(2020)$, were seen, while on the
$(n,M^2)$-trajectories there are only two vacant places. This means
that one state is obviously "superfluous" \  from the point of view
of the $q\bar q$-systematics, i.e. it has to be considered as
exotics. The large value of the width of the $f_2(2000)$ strengthen
the suspicion that, indeed, this state is an exotic one.

\section{Quarkonium and qluonium states: mixing and decay}

On the basis of the $1/N$-expansion rules, we estimate here
effects of mixing of quarkonium and qluonium states. Then, making
use of the rules of quark combinatorics, we give the relations for
decay constants of these states.

\begin{figure}
\centerline{\epsfig{file=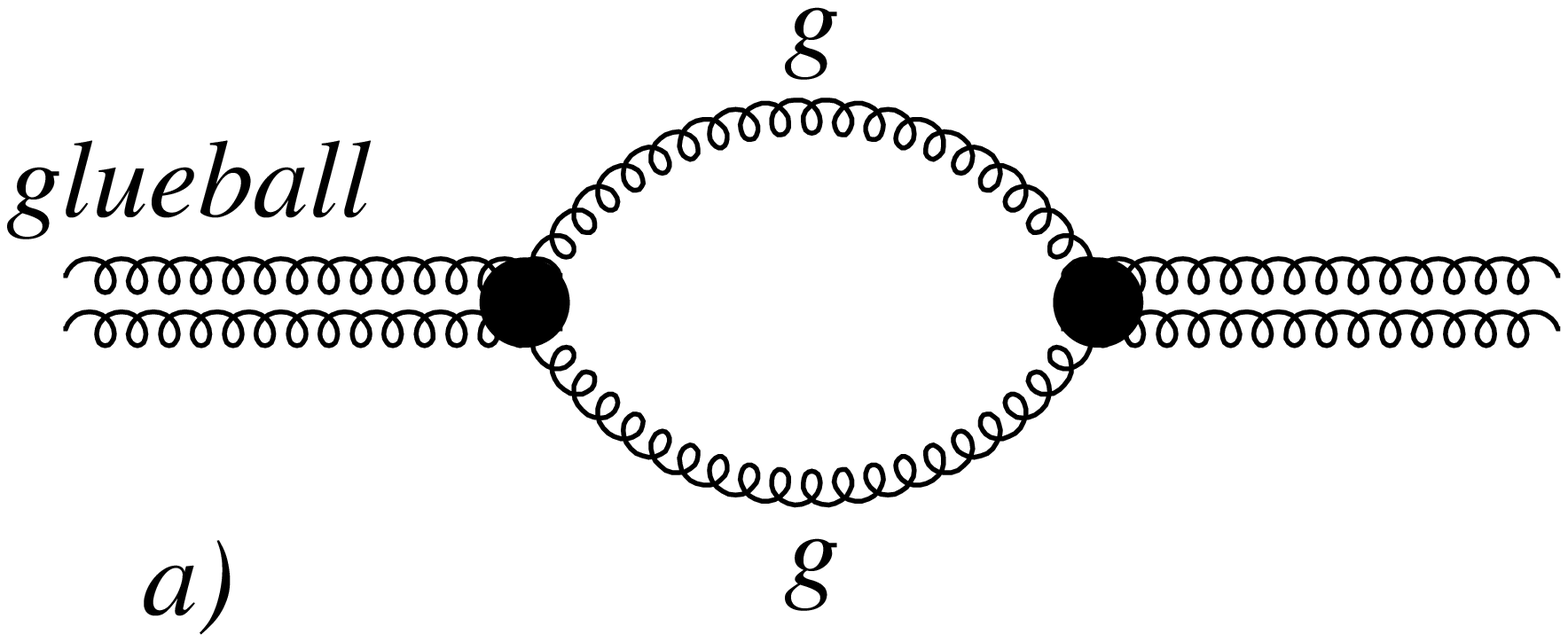,width=8cm}}
\centerline{\epsfig{file=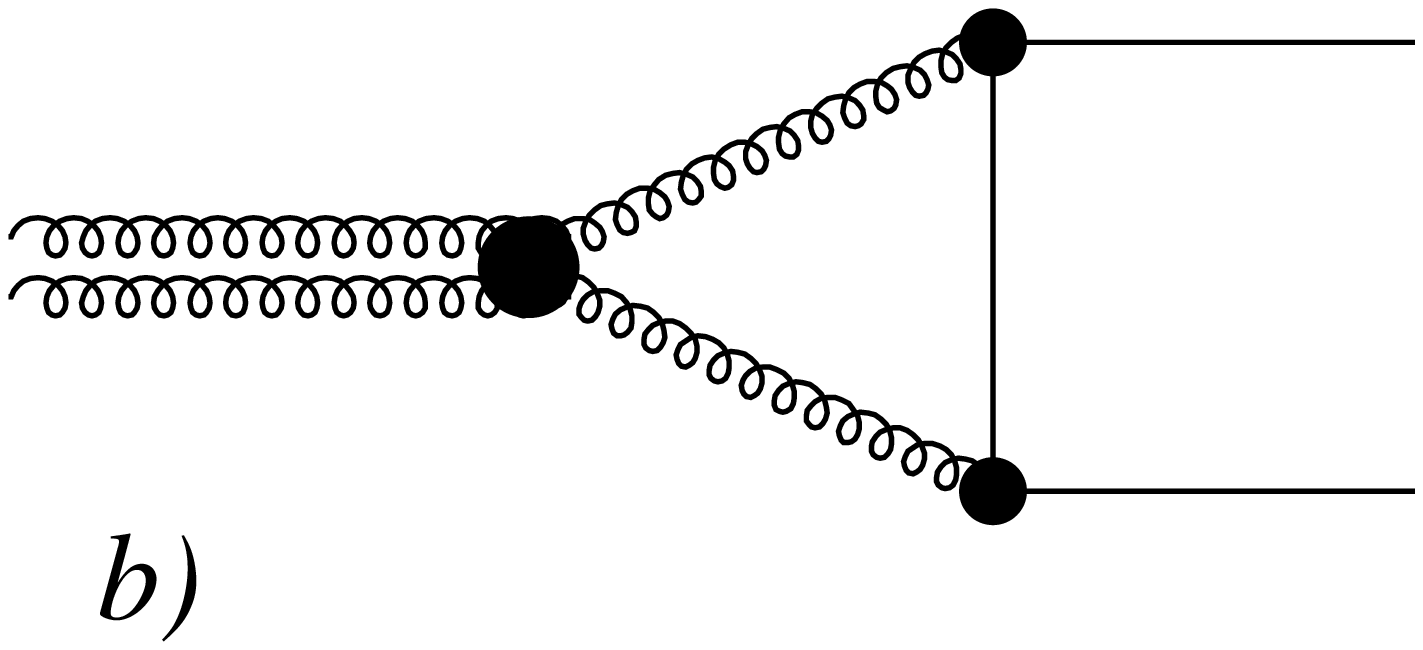,width=8cm}}
\centerline{\epsfig{file=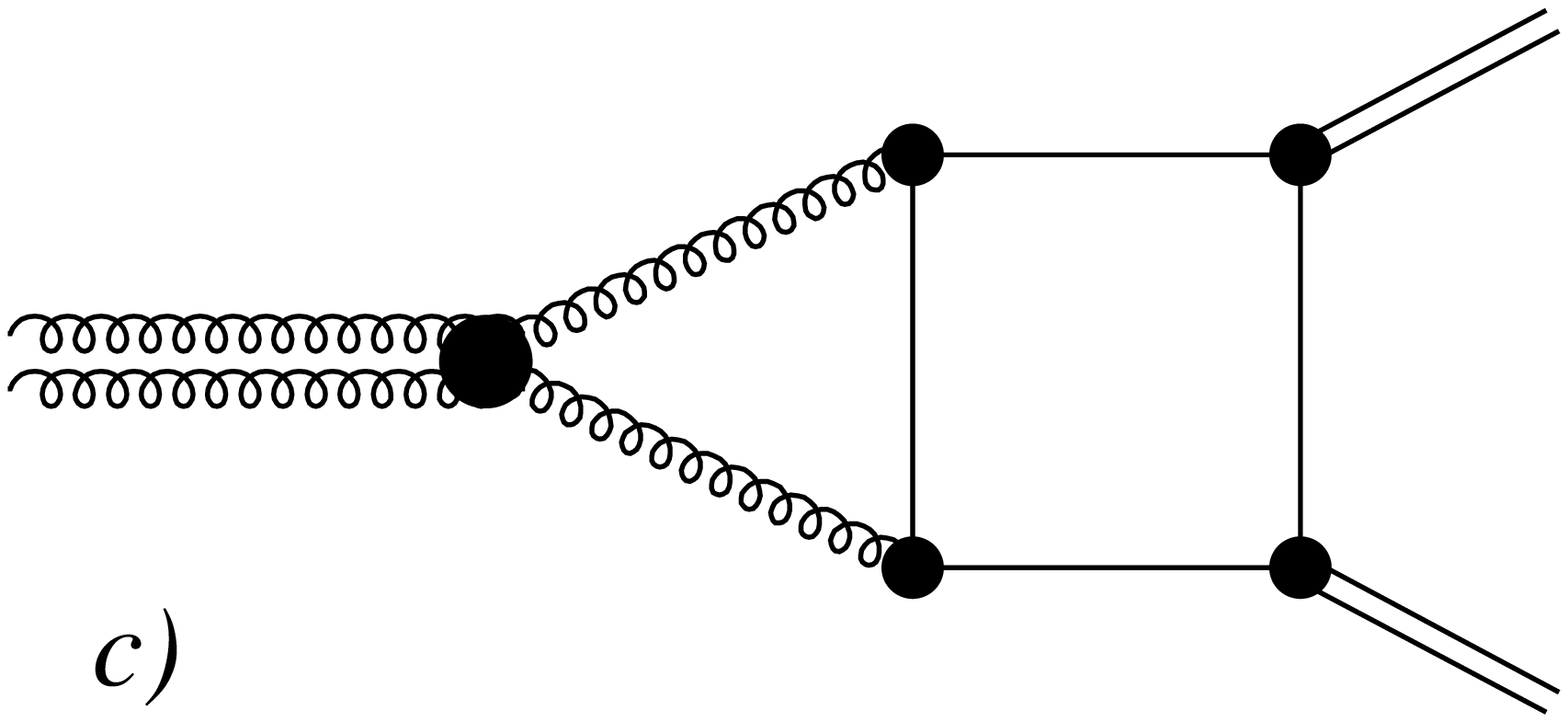,width=8cm}}
\centerline{\epsfig{file=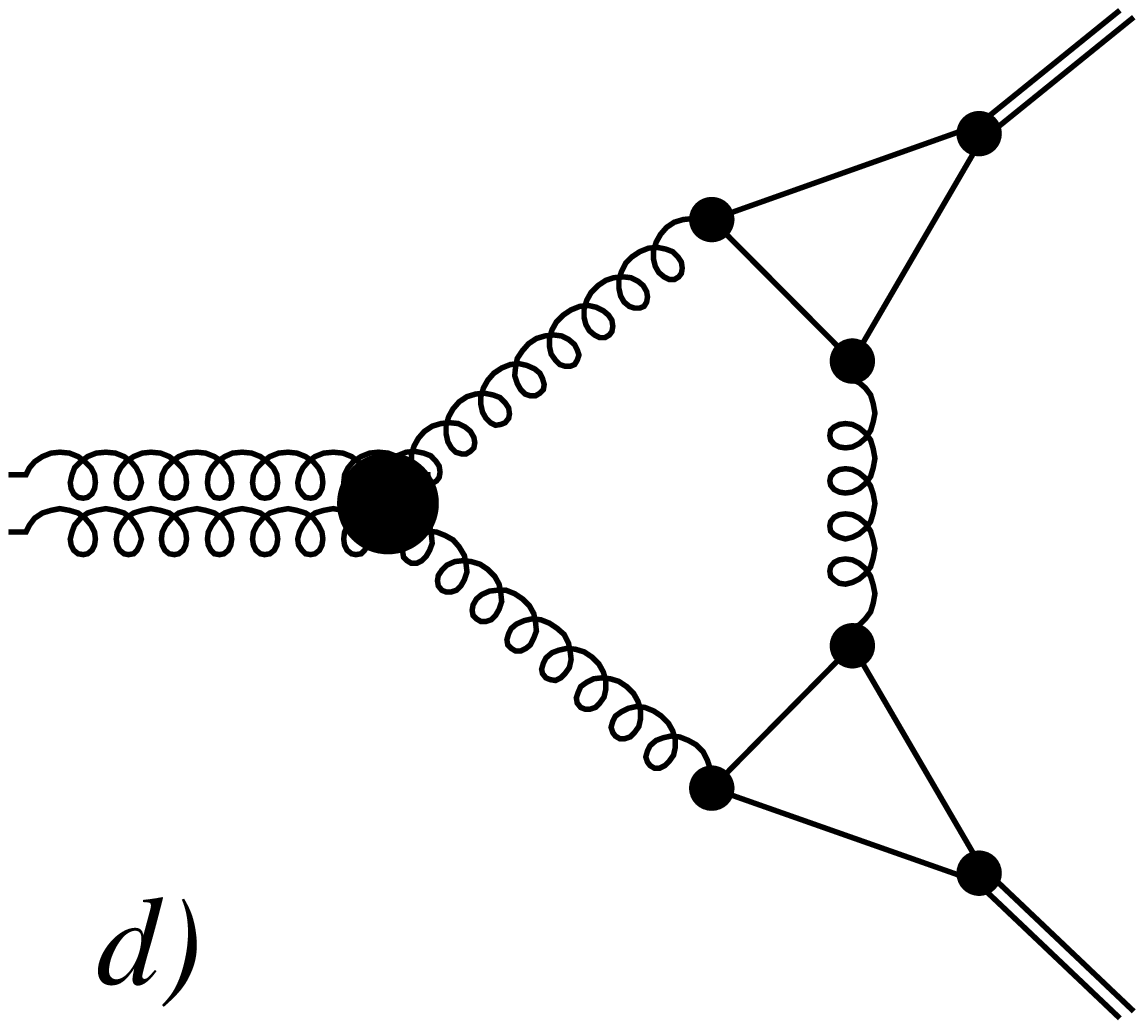,width=8cm}}
\centerline{\epsfig{file=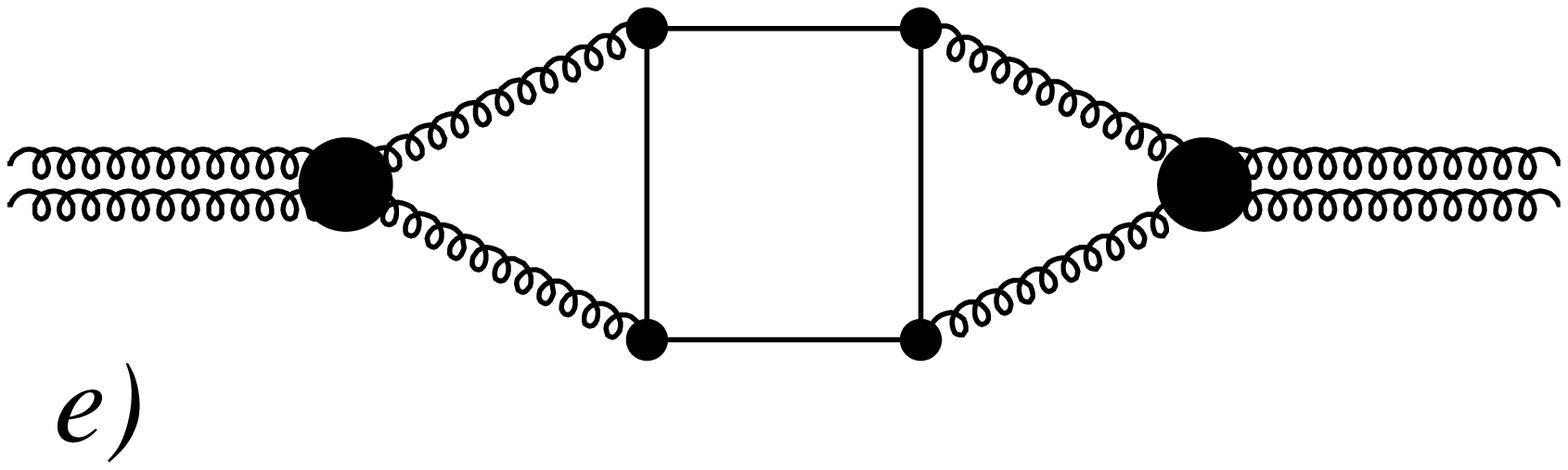,width=8cm}}
\caption{Examles of the diagrams which determine gluonium ($gg$) decay.}
\end{figure}

\begin{figure}
\centerline{\epsfig{file=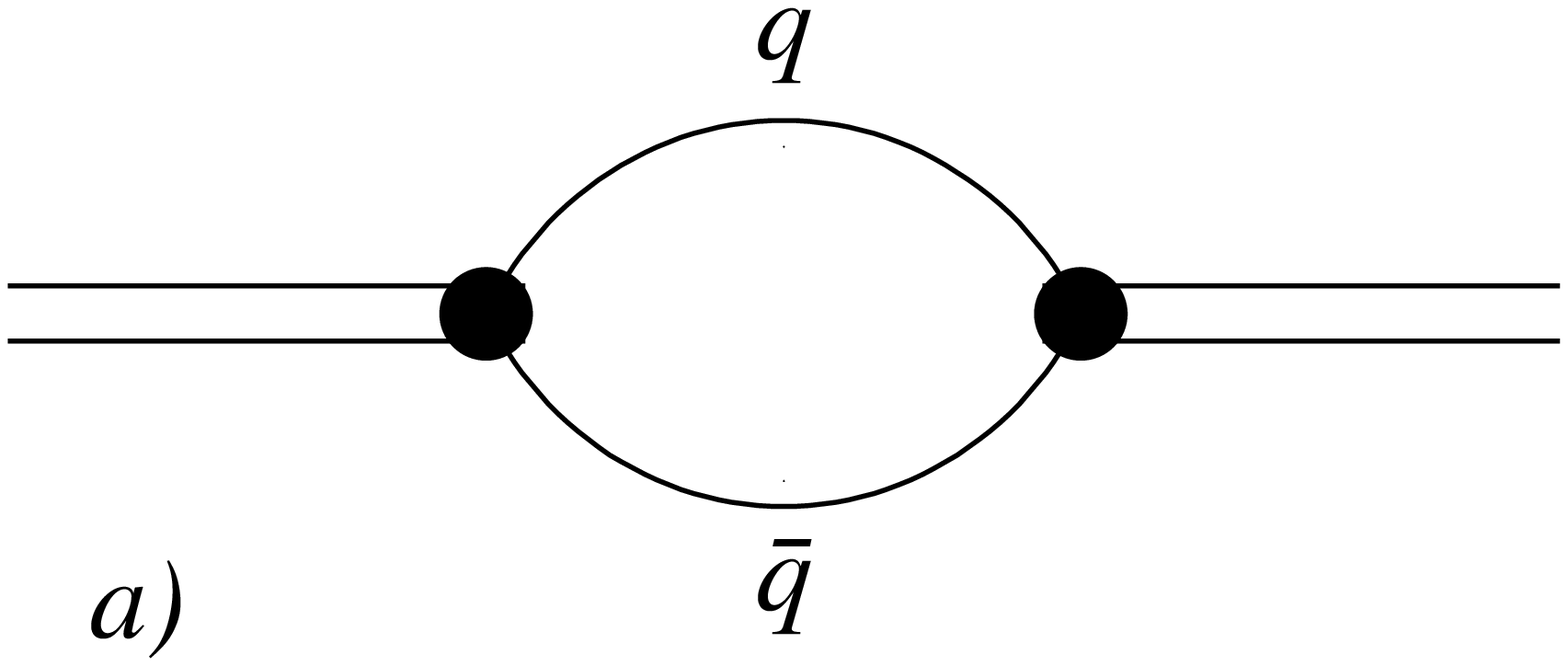,width=8cm}}
\centerline{\epsfig{file=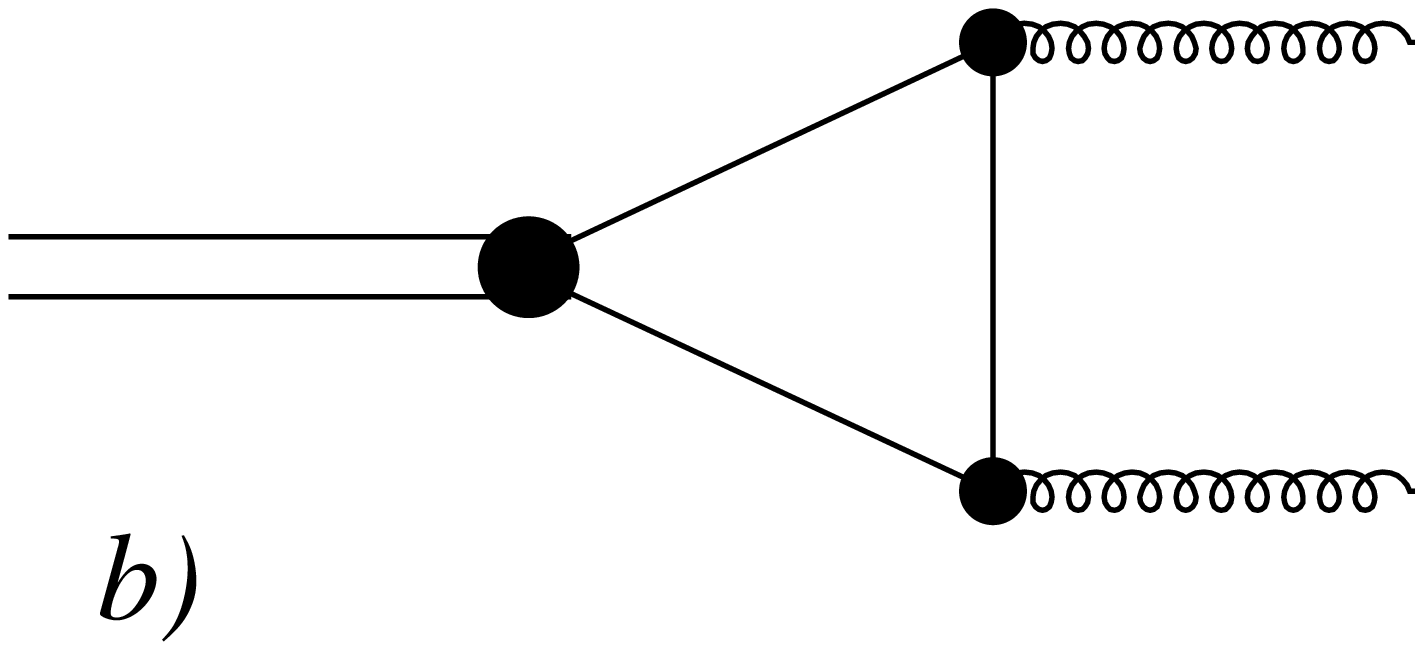,width=8cm}}
\centerline{\epsfig{file=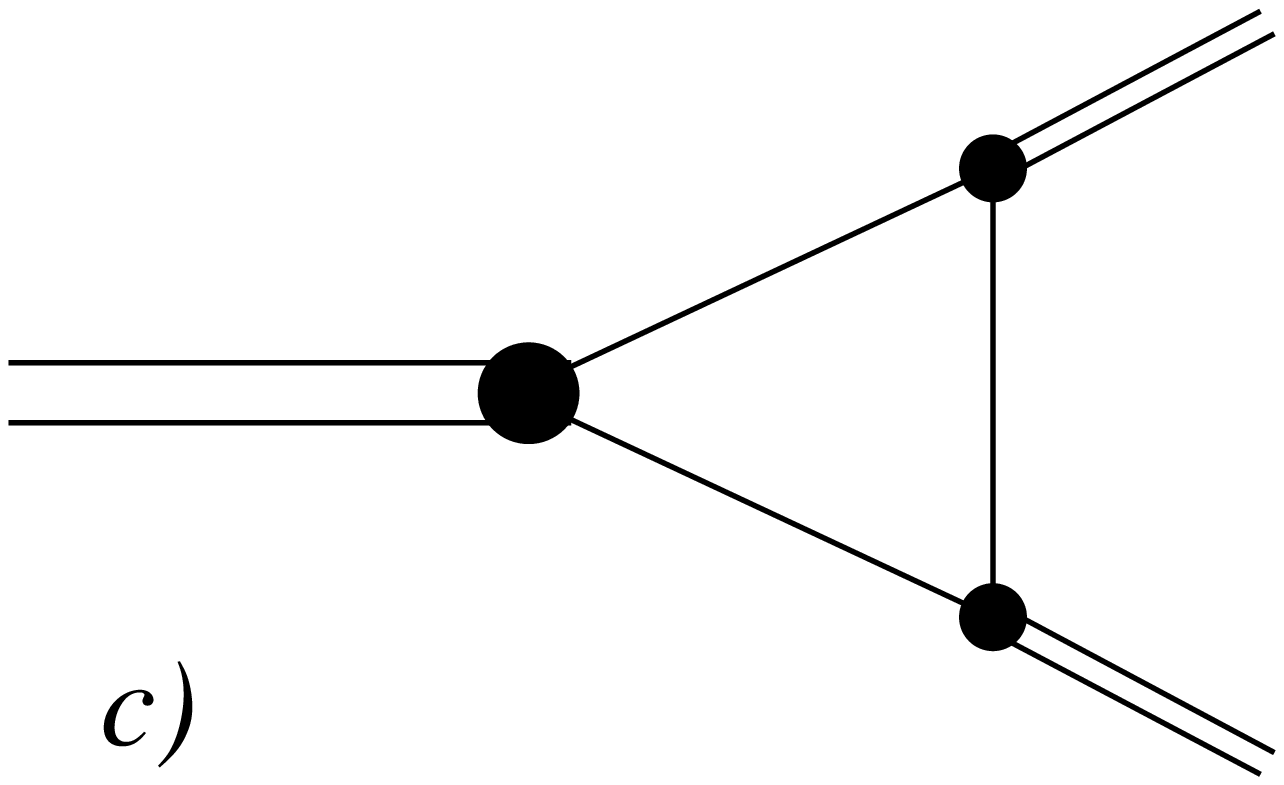,width=8cm}}
\centerline{\epsfig{file=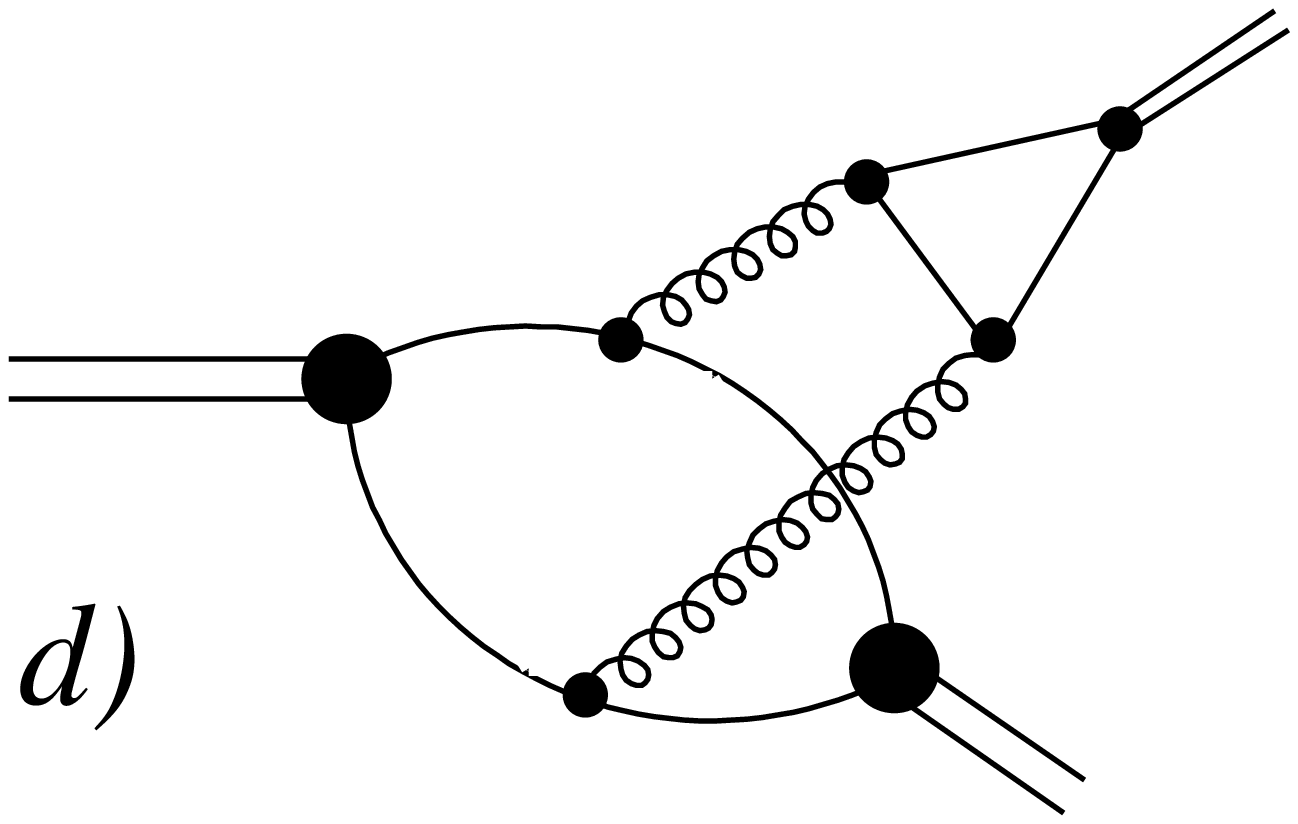,width=8cm}}
\centerline{\epsfig{file=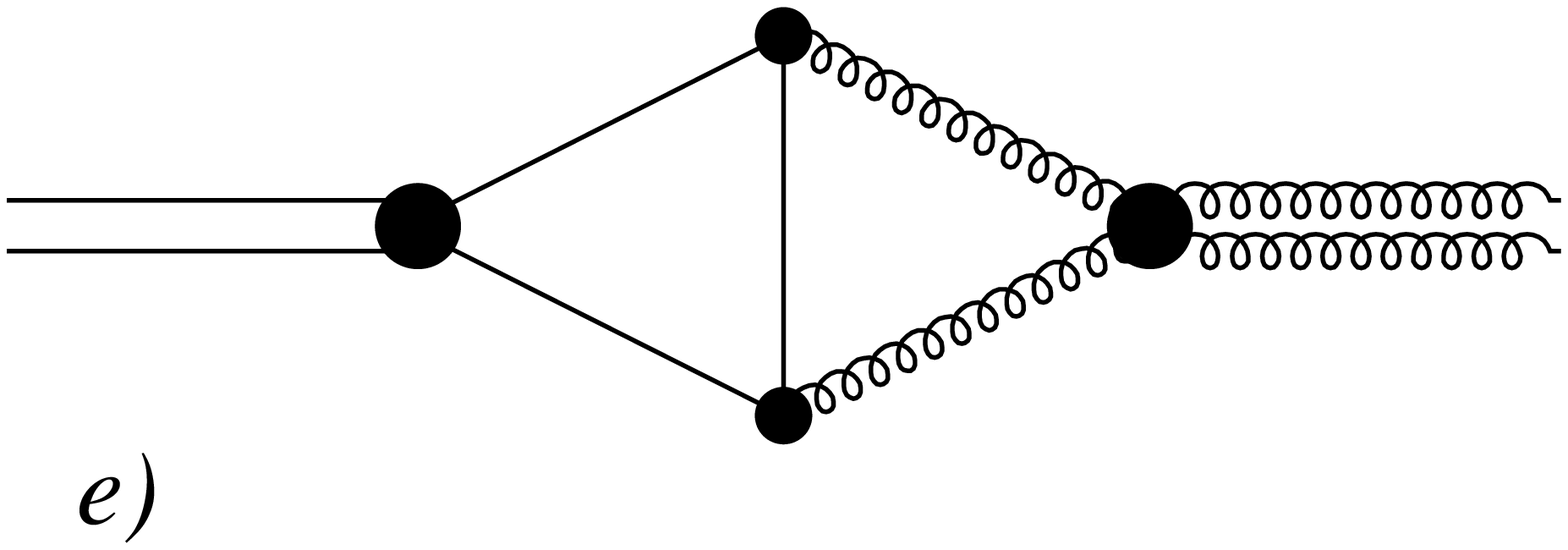,width=8cm}}
\centerline{\epsfig{file=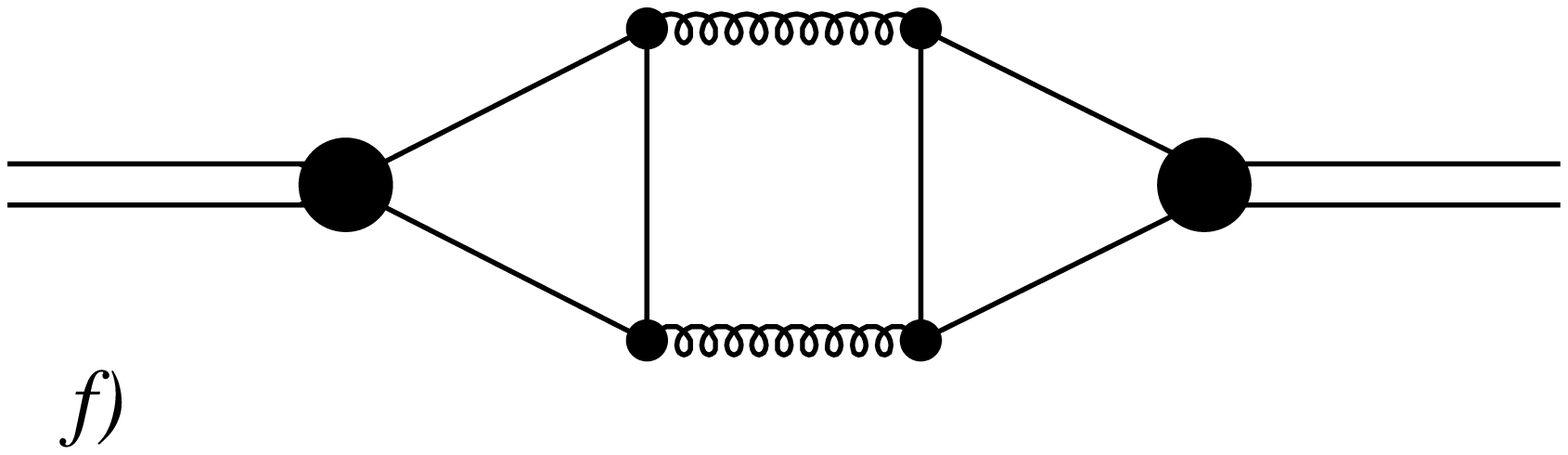,width=8cm}}
\caption{Examles of the diagrams which determine quarkonium ($q\bar q$) decay.}
\end{figure}

\subsection{Mixing of $q\bar q$ and $gg$ states}

The rules of the $1/N$-expansion \cite{t'hooft,ven}, where
$N=N_c=N_f$ are numbers of colours and light flavours, provide a
possibility to estimate the mixing of the gluonium ($gg$) with the
neighbouring quarkonium states ($q\bar q$).

The admixture of the $gg$ component in a $q\bar q$-meson is small,
of the order of $1/N_c$ :
\be \label{3.4}
 f_2(q\bar q-{\rm meson})&=&q\bar q \, \cos
\alpha+gg \,\sin\alpha \\
 \sin^2\alpha &\sim & 1/N_c \, .
\ee
The quarkonium component in the glueball should be larger, it is
of the order of $N_f/N_c$ :
\be \label{3.5}
 f_2({\rm glueball})&=&gg \,\cos
\gamma+(q\bar q)_{glueball}\,\sin\gamma \, , \\
 \sin^2\gamma &\sim & N_f/N_c \, ,
\ee
where $(q\bar q)_{glueball}$ is a mixture of $n\bar n=(u\bar
u+d\bar d)/\sqrt{2}$ and $s\bar s$ components:
\begin{equation} \label{3.6}
 (q\bar q)_{glueball}\ =\ n\bar n\, \cos\varphi_{glueball}+s\bar s \,
\sin\varphi_{glueball}\, ,
\end{equation}
with $\sin\varphi_{glueball}= \sqrt{\lambda/(2+\lambda)}$. If the
flavour SU(3) symmetry were satisfied, the quarkonium component
$(q\bar q)_{glueball}$ would be a flavour singlet,
$\varphi_{glueball}\to \varphi_{singlet} \simeq 37^o$. In reality,
the probability of strange quark production in a gluon field is
suppressed: $u\bar u:d\bar d:s\bar s=1:1:\lambda$, where
$\lambda\simeq 0.5-0.85$. Hence, $(q\bar q)_{glueball}$ differs
slightly from the flavour singlet, it is determined by the
parameter $\lambda$ as follows \cite{Alexei}:
\begin{equation}
(q\bar q)_{glueball}=(u\bar u+d\bar d+\sqrt\lambda\,s\bar
s)/\sqrt{2+ \lambda}\, . \label{3.7}
\end{equation}
The suppression parameter $\lambda$ was estimated both in multiple
hadron production processes \cite{lambda}, and in hadronic decay
processes \cite{klempt,kmat}. In hadronic decays of mesons with
different $J^{PC}$ the value of $\lambda$ can be, in principle,
different. Still, the analyses of the decays of the
$2^{++}$-states \cite{klempt} and $0^{++}$-states \cite{kmat} show
that the suppression parameters are of the same order, 0.5--0.85,
leading to
\begin{equation}
\varphi_{glueball}\ \simeq\ 26^0-33^o.
\label{3.8}
\end{equation}
Let us explain now Eqs. (\ref{3.4})-(\ref{3.7}) in detail.

First, let us evaluate the transition couplings using the rules of
$1/N$-expansion; this evaluation will be done for the decay
transitions {\it gluonium} $\rightarrow$ {\it two $q\bar
q$-mesons} and {\it quarkonium$\rightarrow$ two $q\bar q$-mesons}.
For this purpose, we consider the gluon loop diagram which
corresponds to the two--gluon self--energy part: {\it gluonium}
$\rightarrow$ {\it two gluons} $\rightarrow$ {\it gluonium} (see
Fig. 11a). This loop diagram $ B(gluonium\rightarrow gg\rightarrow
gluonium)$ is of the order of unity, provided the gluonium is a
two--gluon composite system: $ B(gluonium\rightarrow gg\rightarrow
gluonium)\sim g_{gluonium\to gg}^2N_c^2\sim 1$, where
$g_{gluonium\rightarrow gg}$ is a coupling constant for the
transition of a gluonium to two gluons. Therefore,
\begin{equation}
g_{gluonium\to gg}\sim 1/N_c\,.
\label{3.9}
\end{equation}
The coupling constant for the $gluonium\to q\bar q$ transition is
determined by the diagrams of Fig. 11b type. A similar evaluation
gives:
\begin{equation} \label{3.10}
g_{gluonium\rightarrow q\bar q}\sim g_{gluonium\rightarrow gg}\,
g^2N_c\sim 1/N_c\,.
\end{equation}
Here $g$ is the quark--gluon coupling constant, which is of the
order of $1/\sqrt{N_c}$ \cite{t'hooft}. The coupling constant for
the {\it gluonium} $\rightarrow$ {\it two $q\bar q$-mesons}
transition in the leading $1/N_c$ terms is governed by diagrams of
Fig. 11c type:
\begin{equation}
 g_{gluonium\rightarrow two\, mesons}^L\sim
g_{gluonium\rightarrow q\bar q} \, g_{meson\rightarrow q\bar
q}^2N_c\sim 1/N_c\,.
\label{3.11}
\end{equation}
In (\ref{3.11}) the following evaluation of the coupling for
transition $q\bar q -meson\rightarrow q\bar q $ has been used:
 \begin{equation}
 \label{3.12}
g_{meson\rightarrow q\bar q} \sim 1/\sqrt{N_c} \, ,
\end{equation}
which follows from the fact that the loop diagram of the
$q\bar q$-meson propagator (see Fig. 12a) is of the order of
unity: $B(q\bar q-meson\rightarrow q\bar q\rightarrow meson)\sim
g_{meson\rightarrow q\bar q}^2 N_c\sim 1\,.$

The diagram of the type of Fig. 11d governs the couplings for the
transition {\it gluonium} $\rightarrow$ {\it two $q\bar q$-mesons}
in the next-to-leading terms of the $1/N_c$-expansion:
 \begin{equation}
 \label{3.15}
 g_{gluonium\rightarrow two\, mesons}^{NL}\sim g_{gluonium\rightarrow
gg}\,g_{meson\rightarrow gg}^2 N_c^2 \sim 1/N_c^2\, ,
\end{equation}
where the coupling $g_{meson\rightarrow gg}$ has been estimated
following the diagram in Fig. 12b:
 \begin{equation}
 \label{3.14}
g_{meson\rightarrow gg} \sim g_{meson\to q\bar q}\, g^2 \sim
1/N_c^{3/2} .
\end{equation}
Decay couplings of $q\bar q$-meson into two mesons in leading and
next-to-leading terms of $1/N_c$ expansion are determined by
diagrams of the type of Figs. 12c and 12d, respectively. This
gives
 \be \label{3.16}
 g_{meson\to two\, mesons}^{L}
&\sim& g_{meson\rightarrow q\bar q}^3 N_c
\sim 1/\sqrt{N_c}, \\
 g_{meson\to two\, mesons}^{NL}&\sim& g_{meson\rightarrow q\bar q}^2 \,
  g_{meson\rightarrow gg} g^2N_c^2 \sim 1/N_c^{3/2}\,.
\ee
Now we can estimate the order of the value of $\sin^2 \gamma$
which defines the probability $(q\bar q)_ {glueball}$, see Eq.
(\ref{3.5}). This probability is determined by the self-energy
part of the gluon propagator (diagram in Fig. 11e)--- it is of the
order of $N_f/N_c$, the factor $N_f$ being the light flavour
number in the quark loop. Let us emphasise that the diagram in
Fig. 11e stands for only one of the contributions of that type;
indeed, contributions of the same order are also given by diagrams
with all possible (but planar) gluon exchanges in the quark loop.

One can also evaluate $\sin^2 \gamma$ using the transition
amplitude $gluonium \to quarkonium$ (see Fig. 12e), which is of
the order of $1/\sqrt{N_c}$. The value $\sin^2 \gamma$ is
determined by the transition amplitude squared, summed over the
flavours of all quarkonia, thus resulting in Eq. (\ref{3.5}).

The probability of the gluonium component in the quarkonium,
$\sin^2 \alpha$, is of the order of the diagram in Fig. 12f, $\sim
1/N_c$, giving us the estimate (\ref{3.4}). Here, as in the
self-energy gluonium block, planar-type gluon exchanges are
possible. Because of this, in the intermediate $q\bar q$ state all
the interactions are taken into account.

The diagram in Fig. 11e defines also the flavour content of
$(q\bar q)_{glueball}$ --- we see that the gluon field produces
light quark pairs with probabilities $u\bar u:d\bar d:s\bar
s=1:1:\lambda$, so $(q\bar q)_ {glueball}$ is determined by Eq.
(\ref{3.7}) not being a flavour singlet.

\subsection{Quark combinatorial relations for decay constants}

The rules of quark combinatorics lead to relations between decay
couplings for mesons which belong to the same SU(3) nonet. The
violation of the flavour symmetry is taken into account by
introducing a suppression parameter $\lambda$ for the production
of the strange quarks by gluons.

In the leading terms of the $1/N$ expansion, the main contribution
to the decay coupling constant comes from planar diagrams.
Examples of the production of new $q\bar q$-pairs by intermediate
gluons are shown in Figs. 13a and 12b. When an isoscalar
$q\bar q$-meson disintegrates, the coupling constants can be
determined up to a common factor, by two characteristics of a
meson. The first is the quark content of the $q\bar q$-meson,
$ q\bar q=n\bar n\cos\varphi+s\bar s\sin\varphi $, the second is
the parameter $\lambda$. Experimental data provide the following
values for this parameter: $\lambda\simeq 0.5$ \cite{lambda} in
central hadron production in high--energy hadron--hadron
collisions, $\lambda=0.8\pm 0.2$ \cite{klempt} for the decays of
tensor mesons and $\lambda=0.5-0.9$ \cite{kmat} for the decays of
$0^{++}$ mesons.

Let us consider in more detail the production of two pseudoscalar
mesons $P_1P_2$ by $f_2$-quarkonium and $f_2$-gluonium:
\be \label{3.17a}
f_2({\rm quarkonium })& \to&
\pi\pi\,, K\bar K\,,\eta\eta\,,\eta\eta'\,, \eta'\eta'\\
f_2({\rm gluonium })& \to&
\pi\pi\,, K\bar K\,,\eta\eta\,,\eta\eta'\,, \eta'\eta' \, .\nn
\ee
The coupling constants for the decay into channels (\ref{3.17a}),
which in the leading terms of the $1/N$ expansion are
determined by diagrams of the type shown in Fig. 13, may be
presented as
\be \label{3.17}
g^L(q\bar q\to P_1P_2)&=&C^{q\bar q}_{P_1P_2} (\varphi,\lambda)g^L_P\,,
\\
g^L(gg\to P_1P_2)&=&C^{gg}_{P_1P_2}(\lambda)G^L_P\, , \nn
\ee
where
$C^{q\bar q}_{P_1P_2}(\varphi,\lambda)$ and $C^{gg}_{P_1P_2}(\lambda)$
are wholly calculable coefficients depending on the mixing angle
$\varphi$ and parameter $\lambda$; $g^L_P$ and $G^L_P$ are common
factors describing the unknown dynamics of the processes.

Dealing with processes of the Fig. 13b type, one should bear in
mind that they do not contain $(q \bar q)_{quarkonium}$
components in the intermediate state but $(q \bar q)_{continuous\,
spectrum}$ only. The states $(q \bar q)_{quarkonium}$ in this
diagram would lead to processes of Fig. 13c, namely, to a diagram
with the quarkonium decay vertex and the mixing block of $gg$ and
$q\bar q$ components. All these sub-processes are taken into
account separately.

The contributions of the diagrams of the type of Fig. 11d and 12d,
which give the next-to-leading terms, $g^{NL}(q\bar q\to P_1P_2)$ and
$g^{NL}(gg\to P_1P_2)$, may be presented in a form analogous to
(\ref{3.17}). The decay constant to the channel $P_1P_2$ is a sum
of both contributions:
\be \label{3.18}
g^L(q\bar q\to P_1P_2)&+&g^{NL}(q\bar q\to P_1P_2),\\
g^L(gg\to P_1P_2)&+&g^{NL}(gg\to P_1P_2) . \nn
\ee
The second terms are suppressed compared to the first ones by a
factor $N_c$; the experience in the calculation of quark
diagrams teaches us that this suppression is of the order of 1/10.

Coupling constants for gluonium decays, $g^L(gg\to P_1P_2)$ and
$g^{NL}(gg\to P_1P_2)$, are presented in Table 3 while those for
quarkonium decays, $g^L(q\bar q\to P_1P_2)$ and $g^{NL}(q\bar q\to
P_1P_2)$, are given in Table~4.

In Table 5 we give the couplings for decays of the gluonium state
into channels of the vector mesons: $gg\to V_1V_2$.

\begin{figure}[h]
\centerline{\epsfig{file=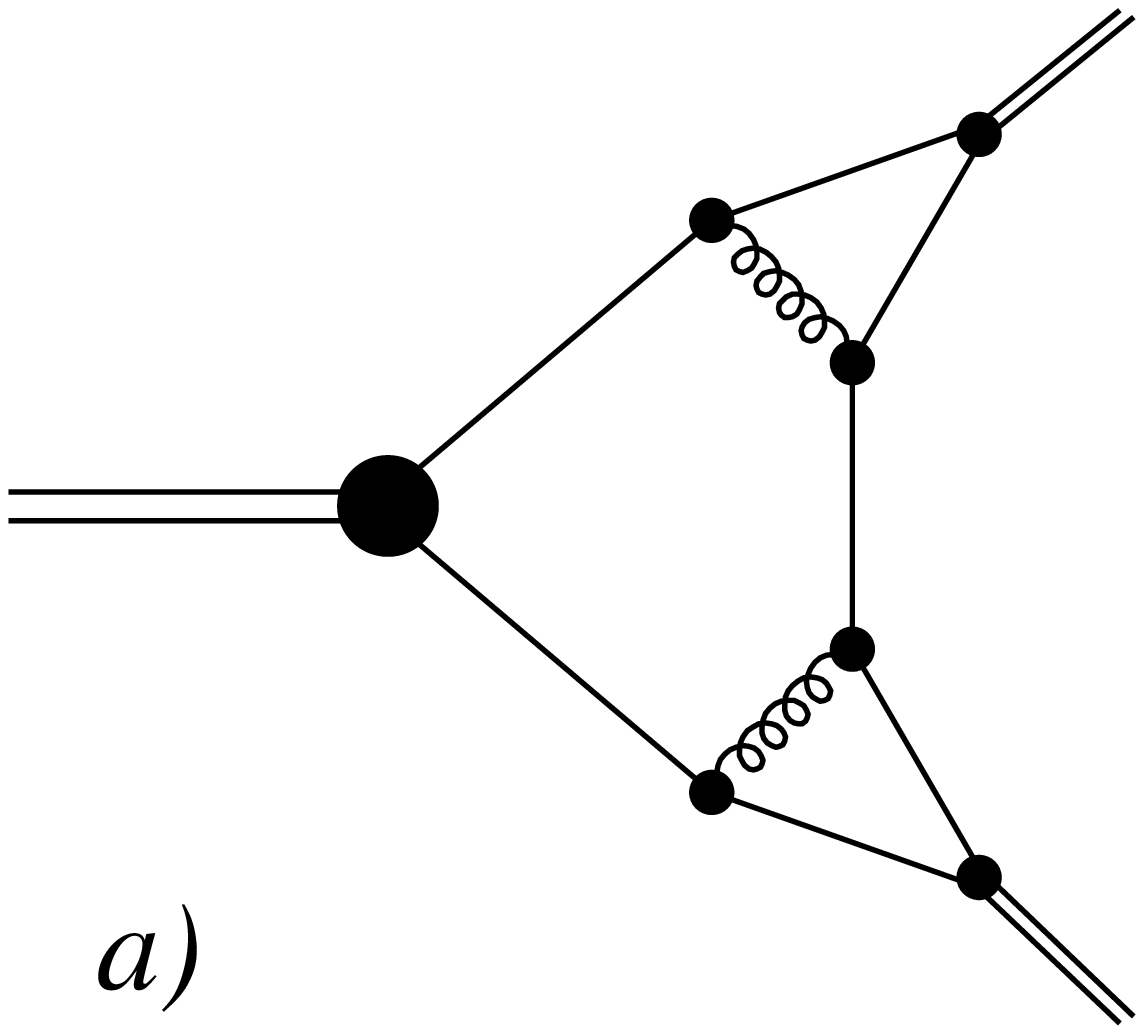,height=5cm}\hspace{0.5cm}
            \epsfig{file=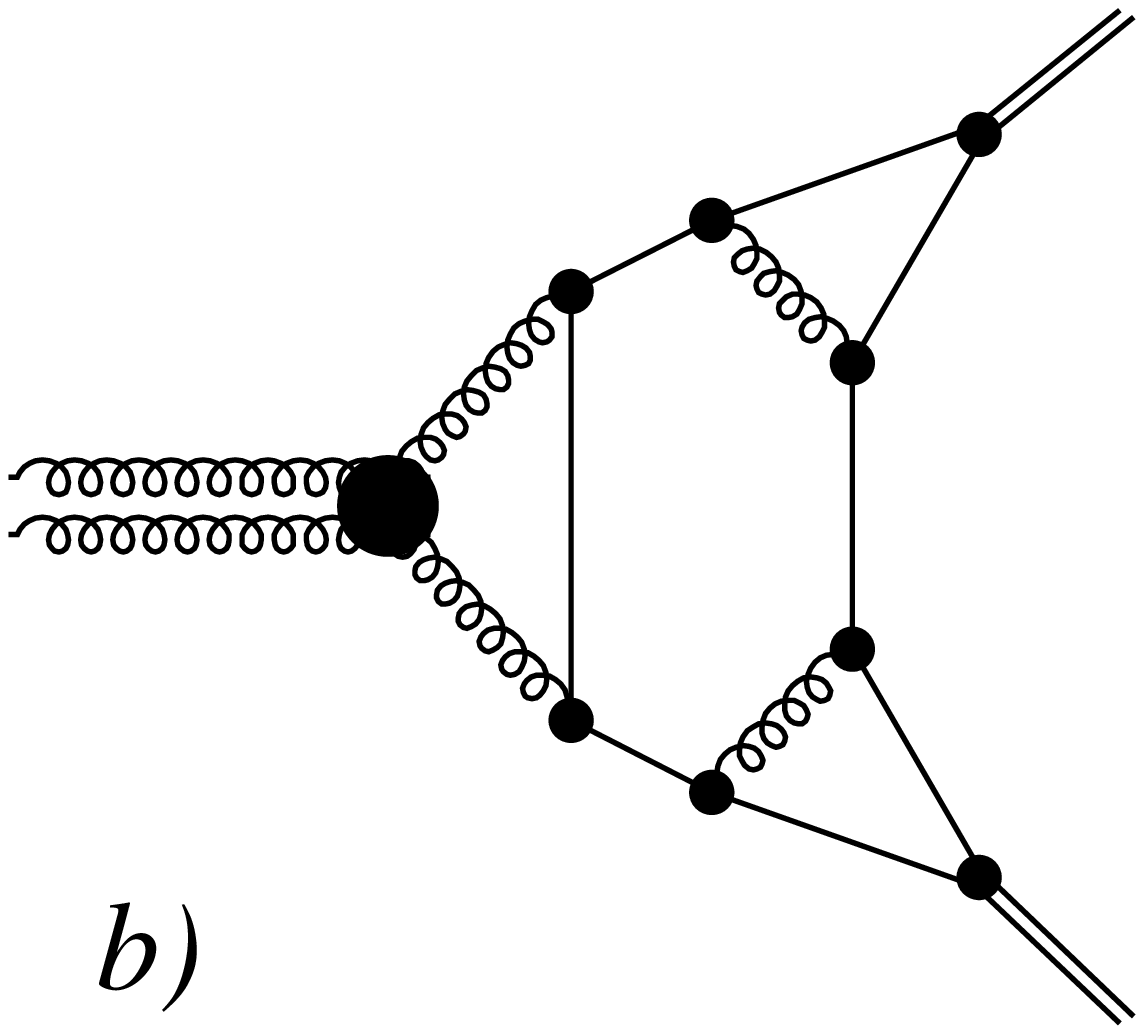,height=5cm}}
\centerline{\epsfig{file=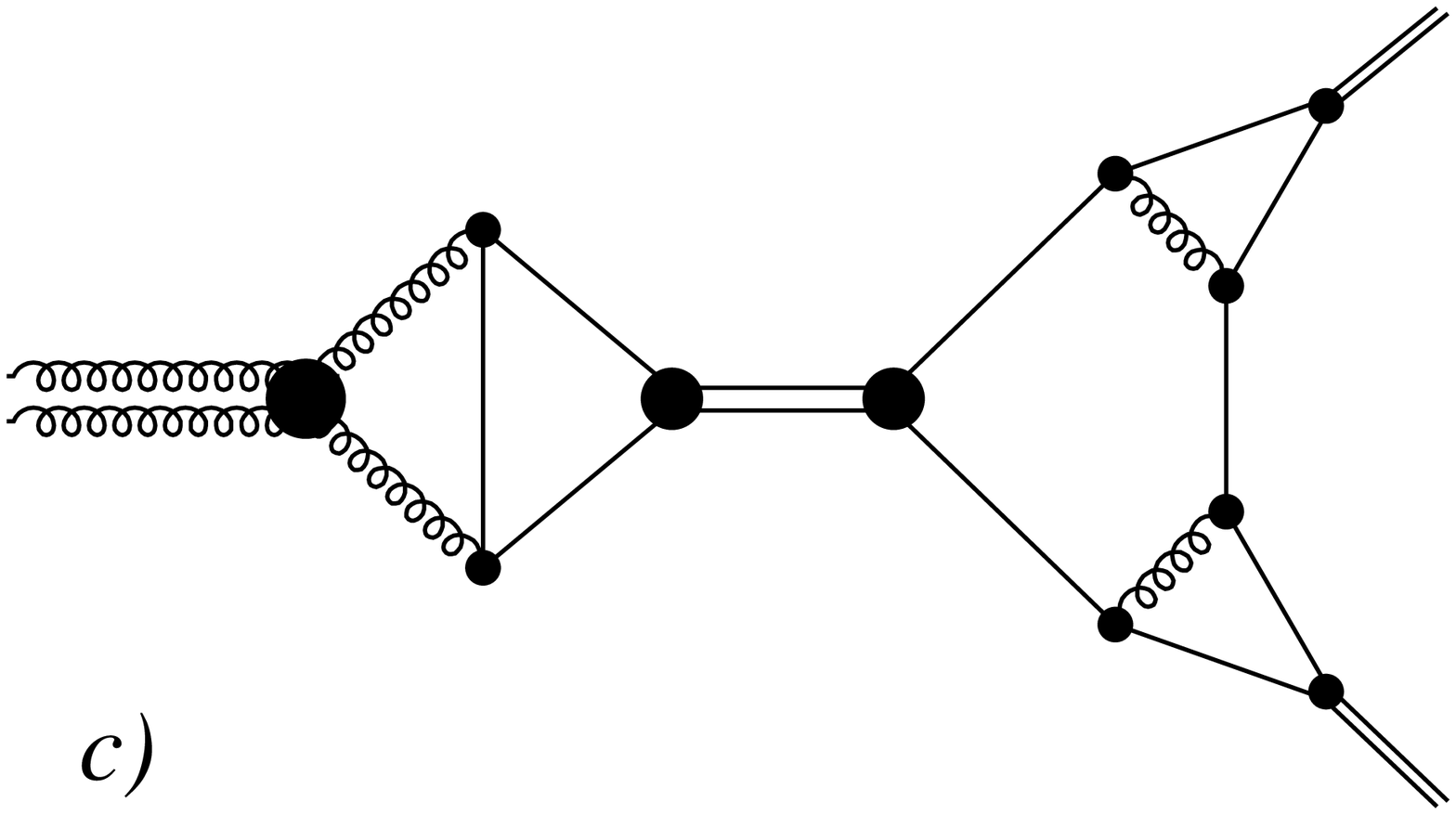,height=5cm}}
\caption{ Examples of planar diagrams responsible for
the decay of the $q\bar q$-state (a) and the gluonium (b) into two
$q\bar q$-mesons (leading terms in the $1/N$ expansion). c) Diagram
for the gluonium decay with a pole in the intermediate
$q\bar q$-state: this process is not included into the gluonium
decay vertex.}
\end{figure}

\newpage
\begin{center}
Table 3\\
Coupling constants of the $f_2$-gluonium decaying to two pseudoscalar
mesons, in the leading\\ and next-to leading terms of $1/N$ expansion.
$\Theta$ is here the mixing angle for $\eta -\eta'$ mesons:
$\eta=n\bar n \cos\Theta-s\bar s \sin \Theta$ and
$\eta'=n\bar n \sin\Theta+s\bar s \cos\Theta$.
\vskip 0.5cm
\begin{tabular}{|c|c|c|c|}
\hline
~ & ~ & ~ &~ \\
~ & Gluonium decay & Gluonium decay &Iden- \\
~ & couplings in the & couplings in the &tity \\
Channel & leading term of &next-to-leading term &factor \\
~ &$1/N$ expansion. &of $1/N$ expansion. & \\ ~
& ~ & ~ &~ \\ \hline ~ & ~ & ~ & ~ \\
$\pi^0\pi^0$ & $G^L$ & 0 & 1/2 \\
~ & ~ & ~ & ~ \\ $\pi^+\pi^-$ & $G^L$ & 0 & 1 \\
~ & ~ & ~ & ~ \\ $K^+K^-$ & $\sqrt \lambda G^L$ & 0 & 1 \\
~ & ~ & ~ & ~ \\ $K^0K^0$ & $\sqrt\lambda G^L $ & 0 & 1 \\
~ & ~ & ~ & ~ \\
$\eta\eta$ & $G^L\left (\cos^2\Theta+ \lambda\sin^2\Theta\right )$
&$2G^{NL}(\cos\Theta-\sqrt{\frac{\lambda}{2}}\sin\Theta )^2$ &
1/2 \\
~ & ~ & ~ & ~ \\
$\eta\eta'$ & $G^L (1-\lambda)\sin\Theta\;\cos\Theta$
&$2G^{NL}(\cos\Theta-\sqrt{\frac{\lambda}{2}}\sin\Theta)\times$ &1 \\
~&~&$(\sin\Theta+\sqrt{\frac{\lambda}{2}}\cos\Theta)$ & ~\\
~ & ~ & ~ & ~ \\
$\eta'\eta'$ &
$G^L\left(\sin^2\Theta+\lambda\;\cos^2\Theta\right)$
&$2G^{NL}\left(\sin\Theta+\sqrt{\frac{\lambda}{2}}\cos\Theta \right)^2$
    & 1/2 \\ ~ & ~ & ~ & ~ \\
\hline
\end{tabular}
\end{center}

\newpage

\begin{center}
Table 4\\
Coupling constants of the $f_2$-quarkonium decaying to two
pseudoscalar mesons in the leading and next-to-leading terms of
the $1/N$ expansion. The flavour content of the $f_2$-quarkonium
is determined by the mixing angle $\varphi$ as follows: $f_2(q\bar
q)=n\bar n \, \cos\varphi +s\bar s\, \sin \varphi$ where $n\bar n
=(u\bar u+d\bar d)/\sqrt 2$.

\vskip 0.5cm

\begin{tabular}{|c|c|c|}
\hline
~ & ~ & ~ \\
~ & Decay couplings of &Decay couplings of \\
~ & quarkonium & quarkonium \\
Channel& in leading term &in next-to-leading term\\ ~
       & of $1/N$ expansion. &of $1/N$ expansion. \\
\hline
~ &~ & ~ \\ $\pi^0\pi^0$ & $g^L\;\cos\varphi/\sqrt{2}$& 0 \\ ~ & ~ & ~
\\ $\pi^+\pi^-$ & $g^L\;\cos\varphi/\sqrt{2}$ & 0 \\ ~ & ~ & ~ \\
$K^+K^-$ & $g^L (\sqrt 2\sin\varphi+\sqrt \lambda\cos\varphi)/\sqrt 8 $
& 0 \\ ~ & ~ & ~ \\ $K^0K^0$ & $g^L (\sqrt 2\sin\varphi+\sqrt
\lambda\cos\varphi)/\sqrt 8 $ & 0 \\ ~ & ~ & ~ \\ $\eta\eta$ &
$g^L\left (\cos^2\Theta\;\cos\varphi/\sqrt 2+\right .$\hfill &$\sqrt 2
g^{NL}(\cos\Theta-\sqrt{\frac{\lambda}{2}}\sin\Theta )\times$\hfill\\
~ &\hfill$\left . \sqrt{\lambda}\;\sin\varphi\;\sin^2\Theta\right )$ &
\hfill$(\cos\varphi\cos\Theta-\sin\varphi\sin\Theta)$ \\
~ & ~ & ~ \\
$\eta\eta'$ &
$g^L\sin\Theta\;\cos\Theta\left(\cos\varphi/\sqrt 2-\right .$\hfill
&$\sqrt{\frac 12}
g^{NL}\left [(\cos\Theta-\sqrt{\frac{\lambda}{2}}\sin\Theta)\times
\right .$\\
~& \hfill $\left .\sqrt{\lambda}\;\sin\varphi\right ) $ &
\hfill$(\cos\varphi\sin\Theta+\sin\varphi\cos\Theta)$\\
~&~&$+(\sin\Theta+\sqrt{\frac{\lambda}{2}}\cos\Theta)\times$\hfill\\
~&~&\hfill$\left .(\cos\varphi\sin\Theta-\sin\varphi\cos\Theta)\right
]$\\ ~ & ~ & ~ \\ $\eta'\eta'$ &
$g^L\left(\sin^2\Theta\;\cos\varphi/\sqrt 2+\right .$\hfill
&$\sqrt 2
g^{NL} (\sin\Theta+\sqrt{\frac{\lambda}{2}}\cos\Theta)\times$\hfill\\
~&\hfill $\left .\sqrt{\lambda}\;\sin\varphi\;\cos^2\Theta\right)$ &
\hfill$(\cos\varphi\cos\Theta+\sin\varphi\sin\Theta )$ \\
~ & ~ & ~ \\
\hline
\end{tabular}
\end{center}

\newpage

\begin{center}
Table 5\\
The constants of the tensor glueball decay into two
vector mesons in the leading (planar diagrams) and next-to-leading
(non-planar diagrams) terms of $1/N$-expansion.
The mixing angle for $\omega-\phi$ mesons is defined as:
$\omega=n\bar n\cos\varphi_V-s\bar s\sin\varphi_V$,
$\phi=n\bar n\sin\varphi_V+s\bar s\cos\varphi_V$. Because of the small
value of $\varphi_V$, we keep in the Table only terms of the order
of $\varphi_V$.

\vskip 0.5cm

\begin{tabular}{||c|c|c|c||}

\hline

 & Constants for & Constants for & Identity factor \\
 & glueball decays in & glueball decays in & for decay \\
Channel & the leading order & next-to-leading order & products \\
 & of $1/N$ expansion & of $1/N$ expansion & \\
\hline

$\rho^0\rho^0$ & $G^L_V$ & 0 & 1/2 \\

$\rho^+\rho^-$ & $G^L_V$ & 0 & 1 \\

$K^{*+}K^{*-}$ & $\sqrt\lambda\,G^L_V$ & 0 & 1 \\

$K^{*0}\bar K^{*0}$ & $\sqrt\lambda\, G^L_V$ & 0 & 1 \\

$\omega\omega$ & $G^L_V$ & $2G^{NL}_V$ & 1/2 \\

$\omega\phi$ & $G^L_V(1-\lambda)\varphi_V$ &

$2G^{NL}_V\left(\sqrt{\frac\lambda2}+\varphi_V

\left(1-\frac\lambda2\right)\right)$ & 1 \\

$\phi\phi$ & $\lambda\,G^L_V$ & $2G^{NL}_V\left(\frac\lambda2

+\sqrt{2\lambda}\,\varphi_V\right)$ & 1/2 \\

\hline
\end{tabular}
\end{center}

 \section{The broad state $f_2(2000)$: the tensor glueball}

In the leading terms of $1/N_c$-expansion we have definite ratios
for the glueball decay couplings. The next-to-leading terms in the
decay couplings give corrections of the order of $1/N_c$.
Let us remind that, as we see in the numerical calculations of the
diagrams, the $1/N_c$ factor leads to a smallness of the order of
$1/10$, and we neglect them in the analysis of the decays
$ f_2\to\pi^0\pi^0, \eta\eta,\eta\eta'$.

Considering a glueball state which is also a mixture of the
gluonium and quarkonium components, we have
$\varphi\to\varphi_{glueball}=\sin^{-1}\sqrt{\lambda/(2+\lambda)}$
for the latter. So we can write
\beq \label{3.18a}
\frac{g^L((q\bar q)_{glueball}\to P_1P_2)}
{g^L((q\bar q)_{glueball}\to P'_1P'_2)} =
\frac{g^L(gg\to P_1P_2)}{g^L(gg\to P'_1P'_2)}
\eeq
Then the relations for decay couplings of the glueball in the
leading terms of the $1/N$-expansion read:
\begin{eqnarray}
&& g^{glueball}_{\pi^0\pi^0}\ =\
\frac{G^L_{glueball}}{\sqrt{2+\lambda}}\ ,
\nonumber\\
&& g^{glueball}_{\eta\eta}\ =\
\frac{G^L_{glueball}}{\sqrt{2+\lambda}}\,
(\cos^2\Theta+\lambda\sin^2\Theta)
\nonumber\\
&& g^{glueball}_{\eta\eta'}\ =\
\frac{G^L_{glueball}}{\sqrt{2+\lambda}}\,
(1-\lambda)\sin\Theta\cos\Theta\ .
\label{4.20}
\end{eqnarray}
Hence, in spite of the unknown quarkonium components in the glueball,
there are definite relations between the couplings of the glueball state
with the channels $\pi^0\pi^0,\eta\eta,\eta\eta'$ which can serve as
signatures to define it.

\subsection{Ratios between coupling constants of \boldmath $f_2(2000)\to
\pi^0\pi^0,\eta\eta,\eta\eta'$ as indication of a glueball nature
of this state}

Eq. (\ref{4.20}) tells us that for the glueball state the relations
between the coupling constants are
$1:(\cos^2\Theta+\lambda\sin^2\Theta):(1-\lambda)\cos\Theta\sin\Theta$.
For $(\lambda=0.5$, $\Theta=37^\circ)$ we have $1:0.82:0.24$, and for
$(\lambda=0.85$, $\Theta=37^\circ)$, respectively, $1:0.95:0.07$.
Consequently, the relations between the coupling constants
$g_{\pi^0\pi^0}:g_{\eta\eta}:g_{\eta\eta'}$ for the glueball have to be
\begin{equation}
\hspace*{-3cm} 2^{++}glueball \hspace{1cm}
g_{\pi^0\pi^0}:g_{\eta\eta}:g_{\eta\eta'}\ =\
1:(0.82-0.95):(0.24-0.07).
\label{4.21}
\end{equation}
It follows from (\ref{2.4}) that only the coupling constants of
the broad $f_2(2000)$ resonance are inside the intervals: $0.82\le
g_{\eta\eta}/ g_{\pi^0\pi^0}\le 0.95$ and $0.24\ge
g_{\eta\eta'}/g_{\pi^0\pi^0}\ge 0.07$. Hence, it is just this
resonance which can be considered as a candidate for a tensor glueball,
while $\lambda$ is fixed in the interval $0.5\le\lambda\le 0.7$. Taking
into account that there is no place for $f_2(2000)$ on the
$(n,M^2)$-trajectories (see Fig.~10), it becomes evident that indeed,
this resonance is the lowest tensor glueball.

\begin{figure}[h]
\centerline{\epsfig{file=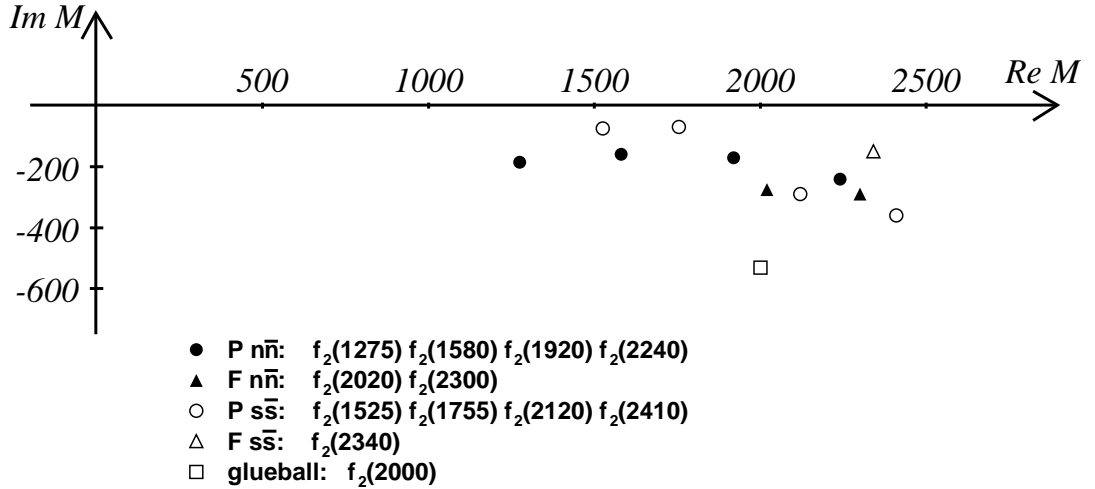,width=15cm}}
\caption{Position of the $f_2$-poles in the complex-$M$ plane: states
with dominant $^3P_2n\bar n$-component (full circle), $^3F_2n\bar
n$-component (full triange), $^3P_2s\bar s$-component (open circle),
$^3F_2s\bar s$-component (open triangle), glueball (open square).}
\end{figure}

\subsection{Mixing of the glueball with neighbouring $q\bar q$-resonances}

The position of the $f_2$-poles on the complex $M$-plane is shown
in Fig. 14. We see that the glueball state $f_2(2000)$ overlaps
with a large group of $q\bar q$-resonances. This means that there
is a considerable mixing with the neighbouring resonances. The
mixing can take place both at relatively small distances, on the
quark-gluon level (processes of the type shown in Fig. 12e), and
owing to decay processes
\beq
f_2(glueball)\to real\,\, mesons\to f_2(q\bar q-meson).
\label{A23}
\eeq
Processes of the type of (\ref{A23}) are presented in Fig. 15.

\begin{figure}[h]
\centerline{\epsfig{file=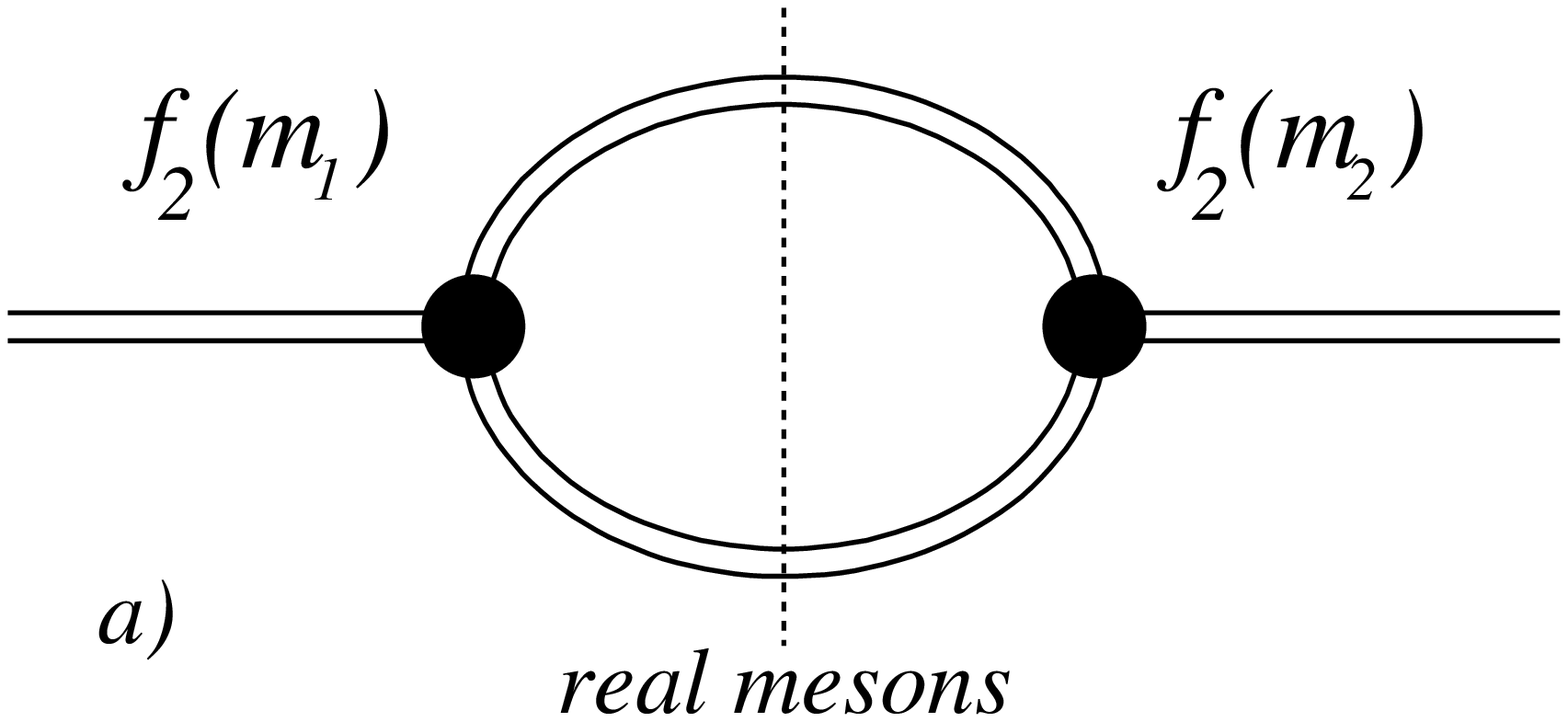,width=8cm}\hspace{0.5cm}
            \epsfig{file=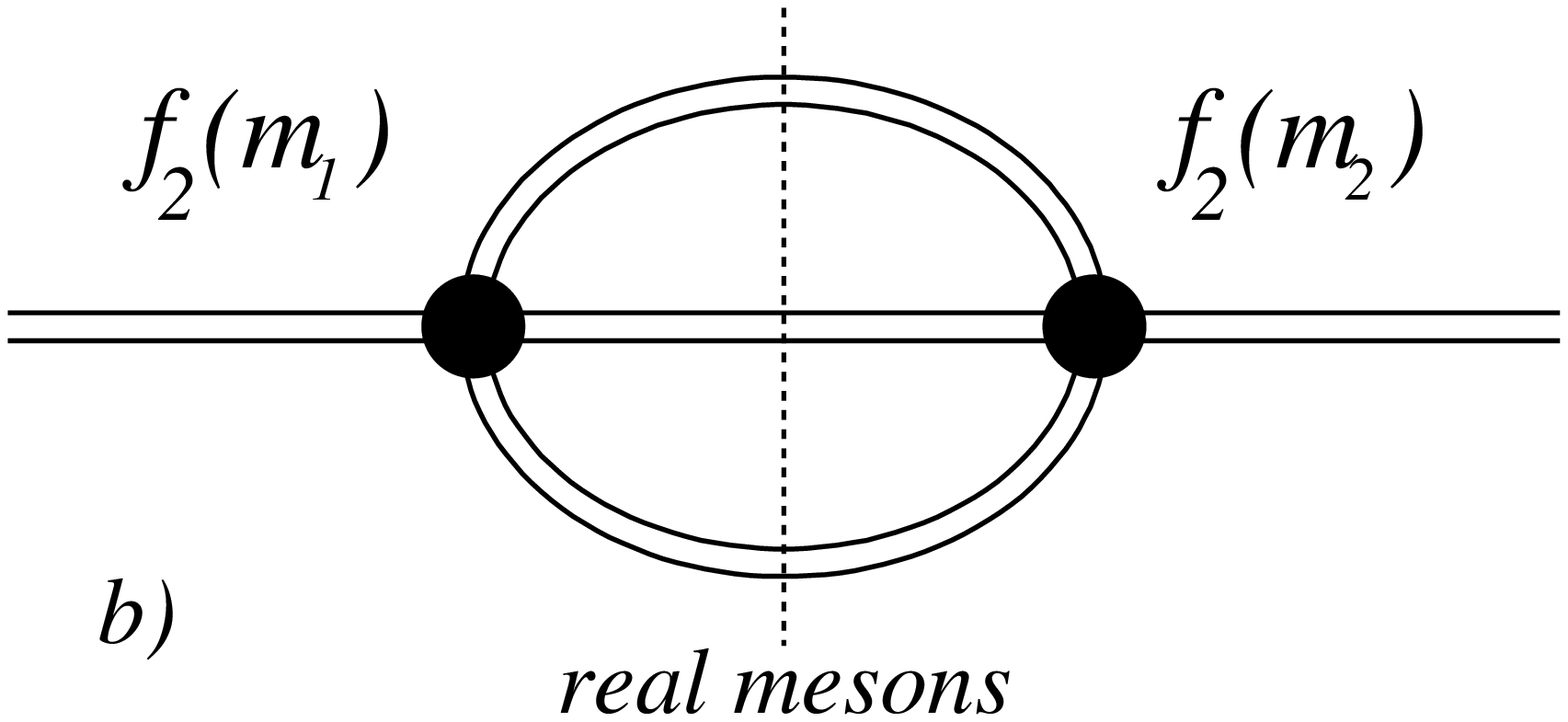,width=8cm}}
\caption{Transitions $f_2(m_1)\to real\,\, mesons\to f_2(m_2)$,
responsible for the accumulation of widths in the case of
overlapping resonances.}
\end{figure}

The estimates which were carried out in Section 4 demonstrated
that even the mixing at the quark-gluon level (diagrams of the
types in Fig. 12e) leads to a sufficiently large admixture of the
quark-antiquark component in the glueball:
$f_2(glueball)=\cos{\gamma gg}+\sin{\gamma gg}$ with
$\sin{\gamma}\sim\sqrt{N_f/N_c}$. A mixing due to processes
(\ref{A23}), apparently, enhances the quark-antiquark component.
The main effect of the processes (\ref{A23}) is, however, that in
the case of overlapping resonances one of them accumulates the
widths of the neighbouring resonances. The position of the
$f_2$-poles in Fig. 14 makes it obvious that such a state is the
tensor glueball.

A similar situation was detected also in the sector of scalar mesons
in the region $1000-1700$\, MeV: the scalar glueball, being in the
neighbourhood of $q\bar q$-resonances, accumulated a relevant
fraction of their widths and transformed into a broad
$f_0(1200-1600)$ state. Such a transformation of a scalar glueball
into a broad state was observed in \cite{ufn04,kmat}; further
investigations verified this observation. The possibility that a
scalar (and tensor) glueball may considerably mix with $q\bar
q$-states was discussed already for quite a long time, see, e.g.,
\cite{ufn97,close,wein}. At the same time there is a number of
papers, e.g. \cite{close,wein}, in which the mixing due to
transitions (\ref{A23}) is not taken into account. Hence, in these
papers relatively narrow resonances like $f_0(1500)$ and $f_0(1710)$
are suggested as possible scalar glueballs.

We see that both glueballs, the scalar $f_0(1200-1600)$ and the
tensor $f_2(2000)$ one, reveal themselves as broad resonances. We
can suppose that this is not accidental. In \cite{AAS-PL} the
transition of the lowest scalar glueball into a broad resonance
was investigated in the framework of modelling decays by
self-energy quark an gluon diagrams. As it was discovered, it
was just the glueball which, appearing among the $q\bar q$-states,
began to mix with them actively, accumulating their widths.
Hence, the glueball turned out to be the broadest resonance.

\subsubsection{The mixing of two unstable states}

In the case of two resonances, the propagator of the state 1
is determined by the diagrams of Fig. 16a. With all these
processes taken into account, the propagator of the state 1
is equal to:
\beq
D_{11}(s)=\left(m_1^2-s-B_{11}(s)-\frac{B_{12}(s)B_{21}(s)}
{m_2^2-s-B_{22}(s)}\right)^{-1}\,.
\label{20}
\eeq
Here $m_1$ and $m_2$ are masses of the input states 1 and 2, and
the loop diagrams $B_{ij}(s)$ are defined by the spectral integral
\beq
B_{ij}(s)
=\int\limits_{4m^2}^{\infty}\frac{d(s')}{\pi}
\frac{g_i(s')g_j(s')\rho(s')}{s'-s-i0}\, , \label{21}
\eeq
where $g_i(s')$ and $g_j(s')$ are vertices and $\rho(s')$ is the
phase space for the intermediate state. It is helpful to introduce
the propagator matrix $D_{ij}$, where the non-diagonal elements
$D_{12}=D_{21}$ correspond to the transitions $1\to 2$ and $2\to
1$ (see Fig. 16b). The matrix reads:
\beq
\hat D=\left|
\begin{array}{ll} D_{11}& D_{12}\\
D_{21}& D_{22}
\end{array}\right|
=\frac{1}{(M_1^2-s)(M_2^2-s)-B_{12}B_{21}} \left|\begin{array}{cc}
M_2^2-s,& B_{12}\\
B_{21},& M_1^2-s
\end{array}\right|\,. \label{22}
\eeq
Here the following notation is used:
\beq
M_i^2=m_i^2-B_{ii}(s)\qquad\qquad i=1,2\;.
\label{23}
\eeq

\begin{figure}[h]
\centerline{\epsfxsize=11cm \epsfbox{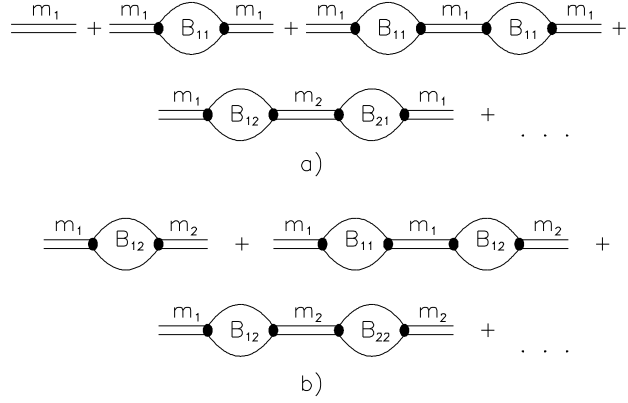}}
\caption{ Diagrams describing the propagation
functions $D_{11}$ (a) and $D_{12}$ (b) for
 the interaction of two resonance states.}
\end{figure}

Zeros in the denominator of the propagator matrix (\ref{22})
define the complex resonance masses after the mixing:
\beq
 \Pi(s)=(M_1^2-s)(M_2^2-s)-B_{12}B_{21}=0\;.
\label{24}
\eeq
Let us denote the complex masses of mixed states as $M_A$ and
$M_B$.

Consider a simple model, where the $s$-dependence of the function
$B_{ij}(s)$ near the points $s\sim M_A^2$ and $s\sim M_B^2$ is
assumed to be negligible. Let $M_i^2$ and $B_{12}$ be constants.
Then one has:
\beq
 M_{A,B}^2=\frac 12 (M_1^2+M_2^2)\pm\sqrt{\frac
14 (M_1^2-M_2^2)^2+ B_{12}B_{21}}\quad.
\label{25}
\eeq
In the case, when the widths of initial resonances 1 and 2 are small
(hence the imaginary part of the transition diagram $B_{12}$ is also
small), the equation (\ref{25}) turns into the standard formula of
quantum mechanics for the split of mixing levels, which
become repulsive as a result of the mixing. If so,
 $$
 \hat D=\left|
\begin{array}{cc}
\cos^2\theta/(M_A^2-s)+sin^2\theta/(M_B^2-s)&
-\cos\theta\sin\theta/(M_A^2-s)+\sin\theta\cos\theta /
(M_B^2-s)\\
-\cos\theta\sin\theta/(M_A^2-s)+\sin\theta\cos\theta /
(M_B^2-s)&
\sin^2\theta/(M_A^2-s)+\cos^2\theta/(M_B^2-s)
\end{array}\right|,
$$

\beq
\cos^2\theta=\frac 12+\frac 12\frac{\frac 12(M_1^2-M_2^2)}
{\sqrt{\frac 14(M_1^2-M_2^2)^2+B_{12}B_{21}}}.
\label{26}
\eeq
The states $|A>$ and $|B>$ are superpositions of the initial
levels, $|1>$ and $|2>$, as follows:
\beq
|A>=\cos\theta|1>-\sin\theta|2>\; , \qquad
|B>=\sin\theta|1>+\cos\theta|2>\; .
\label{27}
\eeq
In general, the representation of states $|A>$ and $|B>$ as
superpositions of initial states is valid, when the $s$-dependence
of functions $B_{ij}(s)$ can not be neglected, and their imaginary
parts are not small. Consider the propagator matrix near $s=M_A^2$:
\beq
\hat D=\frac{1}{\Pi(s)}\left|
\begin{array}{cc} M_2^2(s)-s&B_{12}(s)\\
B_{21}(s)&M_1^2(s)-s
\end{array}\right|
\simeq\,\frac{-1}{\Pi'(M_A^2)(M_A^2-s)} \left|
\begin{array}{cc}
M_2^2(M_A^2)-M_A^2& B_{12}(M_A^2)\\
B_{21}(M_A^2)&
M_1^2(M_A^2)-M_A^2\end{array}\right|\; .
\label{28}
\eeq
In the left-hand side of Eq. (\ref{28}), only the singular (pole)
terms survive. The matrix determinant in the right-hand side of
(\ref{28}) equals zero:
\beq
[M_2^2(M_A^2)-M_A^2][M_1^2(M_A^2)-M_A^2]-B_{12}(M_A^2)B_{21}(M_A^2)=0\quad,
\label{29}
\eeq
This equality follows from Eq. (\ref{24}), which fixes
$\Pi(M_A^2)=0$. It allows us to introduce the complex mixing
angle:
\beq
|A>=\cos\theta_A|1>-\sin\theta_A|2>\,.
\label{30}
\eeq
The right-hand side of Eq. (\ref{26}) can be rewritten by making
use of the mixing angle $\theta_A$, as follows:
\beq
 \left[\hat D\right]_{s\sim
 M_A^2}=\frac{N_A}{M_A^2-s}\left|
\begin{array}{cc}
\cos^2\theta_A&-\cos\theta_A\sin\theta_A\\
-\sin\theta_A\cos\theta_A&\sin^2\theta_A\end{array}\right|\,,
\label{31}
\eeq
where
\beq
N_A=\frac{1}{\Pi'(M_A^2)}[2M_A^2-M_1^2-M_2^2], \,
\cos^2\theta_A=\frac{M_A^2-M_2^2}{2M_A^2-M_1^2-M_2^2}, \,
\sin^2\theta_A=\frac{M_A^2-M_1^2}{2M_A^2-M_1^2-M_2^2}\, .
\label{32}
\eeq
We remind that in the formula (\ref{32}) the functions $M_1^2(s)$,
$M_2^2(s)$ and $B_{12}(s)$ are fixed in the point $s=M_A^2$. In
the case under consideration, when the angle $\theta_A$ is a
complex quantity, the values $\cos^2\theta_A$ and $\sin^2\theta_A$
do not determine the probability of states $|1>$ and $|2>$ in
$|A>$; indeed, the values $\sqrt{N_A}\cos\theta_A$
and $-\sqrt{N_A}\sin\theta_A$ are the transition amplitudes
$|A>\rightarrow |1>$ and $|A>\rightarrow |2>$. Therefore, the
corresponding probabilities are equal to $|\cos\theta_A|^2$ and
$|\sin\theta_A|^2$.

In order to analyse the content of the state $|B>$, an analogous
expansion of the propagator matrix should be carried out near the
point $s=M_B^2$. Introducing
\beq
|B>=\sin\theta_B|1>+\cos\theta_B|2>\,,
\label{33}
\eeq
we have the following expression for $\hat D$ in the vicinity of
the second pole $s=M_B^2$:
\beq
\left[\hat D\right]_{s\sim M_B^2}=\frac{N_B}{M_B^2-s}\left|
\begin{array}{ll}\sin^2\theta_B &\cos\theta_B\sin\theta_B\\
\sin\theta_B\cos\theta_B &\cos^2\theta_B\end{array}\right|\,,
\label{34}
\eeq
where
\beq
N_B=\frac{1}{\Pi'(M_B^2)}\left[2M_B^2-M_1^2-M_2^2\right], \;
\cos^2\theta_B=\frac{M_B^2-M_1^2}{2M_B^2-M_1^2-M_2^2},\;
\sin^2\theta_B=\frac{M_B^2-M_2^2}{2M_B^2-M_1^2-M_2^2}.
\label{35}
\eeq
In Eqs. (\ref{34}), (\ref{35}) the functions $M_1^2(s)$, $M_2^2(s)$ and
$B_{12}(s)$ are fixed in the point $s=M_B^2$.

If $B_{12}$ depends on $s$ weakly and one can neglect this
dependence, the angles $\theta_A$ and $\theta_B$ coincide. In
general, however, they are different. So the formulae for the
propagator matrix differ from the standard approach of quantum
mechanics by this very point.

Another distinction is related to the type of the level shift
afforded by mixing, namely, in quantum mechanics the levels
"repulse" \ each other from the mean value $1/2(E_1+E_2)$ (see also
Eq. (\ref{25})). Generally speaking, the equation (\ref{24}) can
cause both a "repulsion"\  of masses squared from the mean value,
$1/2(M_1^2+M_2^2)$, and an "attraction".

Let us remind now, how to write the amplitudes in the one-channel
and multi-channel cases. The scattering amplitude in the
one-channel case is defined by the following expression:
\beq
 A(s)=g_i(s)D_{ij}(s)g_j(s)\,.
\label{36}
\eeq
In the multi-channel case, $B_{ij}(s)$ is a sum of loop diagrams:
$ B_{ij}(s)=\sum_{a}B_{ij}^{(a)}(s)$, where $B_{ij}^{(a)}$ is a
loop diagram in the channel $a$ with vertex functions $g_i^{(a)}$,
$g_j^{(a)}$ and a phase space factor $\rho_a$. The partial
scattering amplitude in the channel $a\to b$ equals
\beq
A_{a\to b}(s)=g_i^{(a)}(s)D_{ij}(s)g_j^{(b)}(s)\,.
\label{38}
\eeq

\subsubsection{Construction of propagator matrix in a general
case ($N$ resonances)}

Consider the construction of the propagator matrix $\hat D$ for an
arbitrary number $(N)$ of resonances. The matrix elements, $D_{ij}$,
describe the transition from the initial state $i$ (with the bare
propagator $(m_i^2-s)^{-1}$) to the state $j$. They obey the system
of linear equations as follows: \beq
 D_{ij}=D_{ik}B_{kj}(s)(m^2_j-s)^{-1}+\delta_{ij}(m_j^2-s)^{-1}\;,
\label{39}
\eeq
where $B_{ij}(s)$ is the loop diagram for the transition $i \to j$
and $\delta_{ij}$ is the Kronecker symbol.

Let us introduce the diagonal propagator matrix $\hat d$ for
initial states :
\beq
\hat d=diag\left ( (m_1^2-s)^{-1} ,(m_2^2-s)^{-1} ,(m_3^2-s)^{-1}
\cdots \right )\,.
\label{40}
\eeq
Then the system of linear equations (39) can be rewritten in the
matrix form as
\beq
\hat D= \hat D \hat B \hat d +\hat d\;.
\label{41}
\eeq
One obtains
\beq
 \hat D=\frac{I}{(\hat d^{-1}-\hat B)}\;.
\label{42}
\eeq
The matrix $\hat d^{-1}$ is diagonal, hence
$\hat D^{-1}=(\hat d^{-1}-\hat B)$ is of the form
\beq
\hat D^{-1}=\left |
\begin{array}{cccc} M_1^2-s & -B_{12}(s) & -B_{13}(s)
&\cdots\\ -
B_{21}(s) &M_2^2-s & -B_{23}(s) &\cdots\\ -B_{31}(s)
&-B_{32}(s) &M_3^2-s & \cdots\\
\vdots & \vdots & \vdots & \vdots
\end{array}
\right |\;,
\label{43}
\eeq
where $M^2_i$ is defined by Eq. (\ref{23}). Inverting this matrix,
we obtain a full set of elements $D_{ij}(s)$:
\beq
D_{ij}(s)=\frac{(-1)^{i+j}\Pi_{ji}^{(N-1)}(s)}{\Pi^{(N)}(s)}\;.
\label{44}
\eeq
Here $\Pi^{(N)}(s)$ is the determinant of the matrix $\hat D^{-1}$,
and $\Pi_{ji}^{(N-1)}(s)$ is a matrix supplement to the element
$[\hat D^{-1}]_{ji}$, i.e. the matrix
$\hat D^{-1}$ with an excluded $j$-th line and $i$-th column.

The zeros of $\Pi^{(N)}(s)$ define the poles of the
propagator matrix which correspond to physical resonances formed by
the mixing. We denote the complex resonance masses as:
\beq
s=M_A^2\,,\quad M_B^2\,,\quad M_C^2\,, \ldots
\label{45}
\eeq
Near the point $s=M_A^2$, one can leave in
the propagator matrix the leading pole term only.
This means that the free term in Eq. (\ref{41})
can be neglected, so we get a system of homogeneous equations:
\beq
 D_{ik}(s)\left ( \hat d^{-1}-\hat B \right )_{kj}=0\,.
\label{46}
\eeq
The solution of this system is defined up to the
normalisation factor, and it does not depend on the initial index
$i$. If so, the elements of the propagator matrix may be written
in a factorised form as follows:
\beq
 \left[\hat D^{(N)}\right]_{s\sim M_A^2}=\frac{N_A}{M_A^2-s}\cdot
\left|
\begin{array}{llll}\alpha_1^2,&\alpha_1\alpha_2,&
\alpha_1\alpha_3, & \ldots\\
\alpha_2\alpha_1,&\alpha_2^2,&\alpha_2\alpha_3,& \ldots\\
\alpha_3\alpha_1,&\alpha_3\alpha_2,&\alpha_3^2,& \ldots\\
\ldots & \ldots & \ldots & \ldots
\end{array}\right|\,,
\label{47}
\eeq
where $N_A$ is the normalisation factor chosen to satisfy
the condition
\beq
\alpha_1^2+\alpha_2^2+\alpha_3^2+\ldots+\alpha_N^2=1\,.
\label{48}
\eeq
The constants $\alpha_i$ are the normalised amplitudes for the
{\it resonance~A} $\rightarrow$ {\it state} $i$ transitions. The
probability to find the state $i$ in the physical resonance $A$ is
equal to:
\beq
w_i=|\alpha_i|^2\;.
\label{49}
\eeq
Analogous representations of the propagator matrix can be
given also in the vicinity of other poles:
\beq
D_{ij}^{(N)}(s\sim
M_B^2)=N_B\frac{\beta_i\beta_j}{M_B^2-s}\,,\qquad
D_{ij}^{(N)}(s\sim M_C^2)=
N_C\frac{\gamma_i\gamma_j}{M_C^2-s}\, \qquad \cdots.
\label{50}
\eeq
The coupling constants satisfy normalisation conditions similar
to that of Eq. (48):
\beq
 \beta_1^2+\beta_2^2+\ldots+\beta_N^2=1\,,\qquad
\gamma_1^2+\gamma_2^2+\ldots+\gamma_N^2=1\,,\qquad\cdots\,.
\label{51}
\eeq
In the general case, however, there is no completeness condition
for the inverse expansion:
\beq
 \alpha_i^2+\beta_i^2+\gamma_i^2+\ldots\neq 1\,.
\label{52}
\eeq
For two resonances this means that
$\cos^2\Theta_A+\sin^2\Theta_B\neq 1$. Still, let us remind that
the equality in the inverse expansion, which is relevant for the
completeness condition, appears in models where the $s$-dependence
of the loop diagrams is neglected.

\subsubsection{Full resonance overlapping: the accumulation of widths of
neighbouring resonances by one of them}

Let us consider two examples which describe the idealised
situation of a full overlapping of two or three resonances. In
these examples, the effect of accumulation of widths of
neighbouring resonances by one of them can be seen in its original
untouched form.

a) {\it Full overlapping of two resonances.}

For the sake of simplicity, let $B_{ij}$ be a weak $s$-dependent
function; Eq. (\ref{25}) may be used. We define:
\beq
M_1^2=M_R^2-iM_R\Gamma_1\,,\qquad M_2^2=M_R^2-iM_R\Gamma_2\,,
\label{53}
\eeq
and put
\beq
{\rm Re}B_{12}(M_R^2)=P\int\limits_{(\mu_1+\mu_2)^2}^{\infty}
\frac{ds'}{\pi} \frac{g_1(s')g_2(s')\rho(s')}{s'-M_R^2} \to 0\,.
\label{54}
\eeq
It is possible that Re$B_{12}(M_R^2)$ equals zero at positive
$g_1$ and $g_2$, if the contribution from the integration
region $s'<M_R^2$ cancels the contribution from the $s'>M_R^2$
region. In this case
\beq
 B_{12}(M_R^2) \to
ig_1(M_R^2)g_2(M_R^2)\rho(M_R^2)= iM_R\sqrt{\Gamma_1\Gamma_2}\,.
\label{55}
\eeq
Substituting Eqs. (53)--(55) into Eq. (45), one has:
\beq
M_A^2 \to M_R^2-iM_R(\Gamma_1+\Gamma_2)\, \qquad
M_B^2 \to M_R^2\,.
\label{56}
\eeq
Therefore, after mixing, one of the states accumulates the widths of
primary resonances, $\Gamma_A \to \Gamma_1+\Gamma_2$, and another state
becomes a quasi-stable particle, with $\Gamma_B \to 0$.

b) {\it Full overlapping of three resonances.}

Consider the equation
\beq
  \Pi^{(3)}(s)=0
\label{57}
\eeq
in the same approximation as in the above example.
Correspondingly, we put:
\beq
  {\rm Re} B_{ab}(M_R^2) \to 0\, , (a\neq b); \qquad
M_i^2=M_R^2-s-iM_R\Gamma_i=x-i\gamma_i\;.
\label{58}
\eeq
A new variable, $x=M_R^2-s$, is used, and we denote $M_R\Gamma_i=
\gamma_i$. Taking into account $B_{ij}B_{ji}=-\gamma_i\gamma_j$
and $B_{12}B_{23}B_{31}=-i\gamma_1\gamma_2\gamma_3$, we can
rewrite the equation (\ref{57}) as follows:
\beq
x^3+x^2(i\gamma_1+i\gamma_2+i\gamma_3)=0\,.
\label{59}
\eeq
Therefore, at full resonance overlapping, one obtains:
\beq
M_A^2 \to M_R^2-iM_R(\Gamma_1+\Gamma_2+\Gamma_3)\; , \qquad
M_B^2\to M_R^2\; , \qquad M_C^2\to M_R^2\,.
\label{60}
\eeq
Thus, the resonance $A$ has accumulated the widths of three
primary resonances, and the states $B$ and $C$ became
quasi-stable and degenerate.

\section{The \boldmath $q\bar q$-$gg$ content of $f_2$-mesons,
observed in the reactions $p\bar
p\to\pi^0\pi^0,\eta\eta,\eta\eta'$}

We determine here the $q\bar q-gg$ content of $f_2$-mesons,
observed in the reactions $p\bar p\to\pi^0\pi^0,
\eta\eta,\eta\eta'$ \cite{Ani}. This determination is based on
experimentally observed relations (\ref{2.4}) and the rules of
quark combinatorics taken into account in the leading terms of the
$1/N$-expansion.

For the $f_2\to\pi^0\pi^0,\eta\eta,\eta\eta'$ transitions, when
the $q\bar q$-meson is a mixture of quarkonium and gluonium
components, the decay vertices read in the leading terms of the
$1/N$-expansion (see Tables 3 and 4) as follows:
\begin{eqnarray} &&
g^{q\bar q-meson}_{\pi^0\pi^0}\ =\ g\,\frac{\cos\varphi}{\sqrt2}+\frac
G{\sqrt{2+\lambda}}\ , \nonumber\\ &&
g^{q\bar q-meson}_{\eta\eta}=
g\left(\cos^2\theta\frac{\cos\varphi}{\sqrt2}+\sin^2\Theta
\sqrt\lambda \sin\varphi\right)+\frac G{\sqrt{2+\lambda}}
(\cos^2\Theta+\lambda\sin^2\Theta)\ ,
\nonumber\\
&& g^{q\bar q-meson}_{\eta\eta'}=\sin\Theta\cos\Theta\left[g\left(
\frac{\cos\varphi}{\sqrt2}-\sqrt\lambda\sin\varphi\right)+
\frac G{\sqrt{2+\lambda}}(1-\lambda)\right].
\label{4.19}
\end{eqnarray}
The terms proportional to $g$ stand for the $q\bar q\to two\,mesons$
transitions ($g=g^L\cos\alpha$), while the terms with $G$ represent
the $gluonium\to two\,mesons$ transition ($G=G^L\sin\alpha$).
Consequently, $G^2$ and $g^2$ are proportional to the probabilities
for finding gluonium ($W=\sin^2\alpha$) and quarkonium $(1-W)$
components in the considered $f_2$-meson.
Let us remind that the mixing angle $\Theta$ stands for the
$n\bar n$ and $s\bar s$ components in the $\eta$ and $\eta'$
mesons; we neglect the possible admixture of a gluonium component
to $\eta$ and $\eta'$ (according to \cite{eta-glue}, the gluonium
admixture to $\eta$ is less than 5\%, to $\eta'$ --- less than 20\%).
For the mixing angle $\Theta$ we take $\Theta=37^\circ$.

\subsection{The analysis of the quarkonium-gluonium contents of
\newline \boldmath the $f_2(1920)$, $f_2(2020)$, $f_2(2240)$, $f_2(2300)$}

Making use of the data (\ref{2.4}), the relations (\ref{4.19})
allow us to to find $\varphi$ as a function of the ratio $G/g$ of
the coupling constants. The result for the resonances $f_2(1920)$,
$f_2(2020)$, $f_2(2240)$, $f_2(2300)$ is shown in Fig.~17. Solid curves
enclose the values of $g_{\eta\eta}/g_{\pi^0\pi^0}$ for
$\lambda=0.6$ (this is the $\eta\eta$-zone in the $(G/g,\varphi)$
plane) and dashed curves enclose
$g_{\eta\eta'}/g_{\pi^0\pi^0}$ for $\lambda=0.6$ (the
$\eta\eta'$-zone). The values of $G/g$ and $\varphi$, lying in both
zones, describe the experimental data (\ref{2.4}): these regions
are shadowed in Fig.~17.

The correlation curves in Fig.~17 enable us to give a qualitative
estimate for the change of the angle $\varphi$ (i.e. the relation of
the $n\bar n$ and $s\bar s$ components in the $f_2$ meson) depending
on the value of the gluonium admixture. The values $g^2$ and $G^2$
are proportional to the probabilities of having quarkonium and
gluonium components in the $f_2$ meson, $g^2=(g^L)^2(1-W)$ and
$G^2=(G^L)^2W$. Here $W$ is the probability of a gluonium admixture
in the considered $q\bar q$-meson; $g^L$ and $G^L$ are universal
constants, see Tables~3 and 4. Since $G^L/g^L\sim1/\sqrt{N_c}$ and
$W\sim 1/N_c$, we can give a rough estimate:
\begin{equation}
\frac{G^2}{g^2}\ \sim\ \frac W{N_c(1-W)}\to \frac{W}{10}\ .
 \label{4.22}
\end{equation}
Let us remind that the numerical calculations of the diagrams
indicate that $1/N_c$ leads to a smallness of the order of $1/10$ --
this is taken into account in (\ref{4.22}). Assuming that the
gluonium components are less than 20\% ($W<0.2$) in each of the
$q\bar q$ resonances $f_2(1920)$, $f_2(2020)$, $f_2(2240)$,
$f_2(2300)$, we put roughly $W\simeq10\,G^2/g^2$, and obtain for the
angles $\varphi$ the following intervals:
\begin{eqnarray}
&&
W_{gluonium} [f_2(1920)]<20\%: \quad-0.8^\circ<\varphi[f_2(1920)]<
3.6^\circ\ , \nonumber\\
&& W_{gluonium}[f_2(2020)]<20\% :
\quad-7.5^\circ<\varphi[f_2(2020)]< 13.2^\circ\ , \nonumber\\
&& W_{gluonium}[f_2(2240)]<20\%:\quad-8.3^\circ<\varphi[f_2(2240)]
<17.3^\circ\ , \nonumber\\
&&W_{gluonium}[f_2(2300)]<20\% : \quad -25.6^\circ
<\varphi[f_2(2300)] < 9.3^\circ
\label{4.23}
\end{eqnarray}

\begin{figure}[h]
\centerline{\epsfig{file=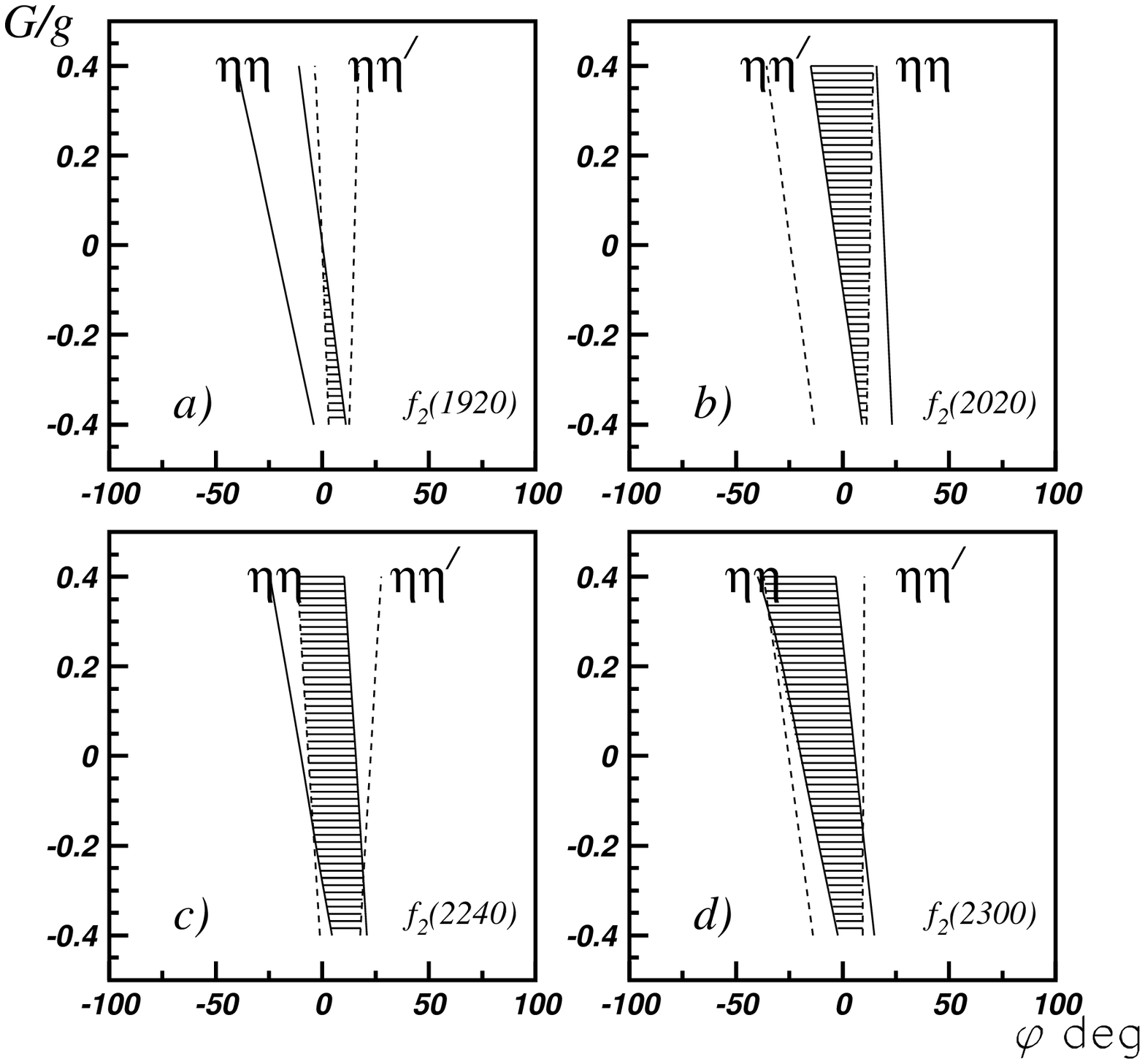,width=15cm}}
\caption{Correlation curves $g_{\eta\eta}(\varphi,G/g)
/g_{\pi^0\pi^0}(\varphi,G/g)$ and
$g_{\eta\eta'}(\varphi,G/g)/g_{\pi^0\pi^0}(\varphi,G/g)$ drawn
according to (64) at $\lambda=0.6$ for
$f_2(1920)$, $f_2(2020)$, $f_2(2240)$, $f_2(2300)$ [2,17].
Solid and dashed curves enclose the values
$g_{\eta\eta}(\varphi,G/g)/g_{\pi^0\pi^0}(\varphi,G/g)$ and
$g_{\eta\eta'}(\varphi,G/g)/g_{\pi^0\pi^0}(\varphi,G/g)$
which obey (3) (the zones $\eta\eta$ and $\eta\eta'$ in the
$(G/g,\varphi)$ plane).
The values of $G/g$ and $\varphi$, lying in both
zones describe the experimental data (3): these are the shadowed
regions. }
\end{figure}

\subsection{\boldmath The $n\bar n$-$s\bar s$ content of the $q\bar
q$-mesons}

Let us summarise what we know about the status of the $(I=0, J^{PC}=
2^{++})$ $q\bar q$-mesons. Estimating the $n\bar n$-$s\bar s$
content of the $f_2$-mesons, we ignore the $gg$ admixture
(remembering that it is of the order of $\sin^2 \alpha \sim 1/N_c$).

\begin{enumerate}

\item The resonances $f_2(1270)$ and $f'_2(1525)$
 are well-known partners of the basic nonet with $n=1$ and a dominant
 $P$-component, $1\,^3P_2q\bar q$. Their flavour content, obtained
from the reaction $\gamma\gamma \to K_SK_S $, is

\begin{eqnarray}
f_2(1270) &=&
\cos\varphi_{n=1}n\bar n+\sin\varphi_{n=1}s\bar s, \nonumber\\
f_2(1525) &=& -\sin\varphi_{n=1}n\bar n+\cos\varphi_{n=1}s\bar s,
\nonumber\\
&& \varphi_{n=1}\ =\ -1\pm 3^\circ.
\label{5.25}
\end{eqnarray}

\item The resonances $f_2(1560)$ and $f_2(1750)$ are partners in a
nonet with $n=2$ and a dominant $P$-component, $2\,^3P_2q\bar q$.
Their flavour content, obtained from the reaction
$\gamma\gamma \to K_SK_S $, is

\begin{eqnarray}
f_2(1560) &=&
\cos\varphi_{n=2}n\bar n+\sin\varphi_{n=2}s\bar s, \nonumber\\
f_2(1750) &=& -\sin\varphi_{n=2}n\bar n+\cos\varphi_{n=2}s\bar s,
\nonumber\\
&& \varphi_{n=1}\ =\ -10^{+5}_{-10}\,^\circ.
\label{5.26}
\end{eqnarray}

\item The resonances $f_2(1920)$ and $f_2(2120)$ \cite{LL} (in \cite{PDG}
they are denoted as $f_2(1910)$ and $f_2(2010)$) are partners in a nonet
with $n=3$ and with a dominant $P$-component, $3\,^3P_2q\bar q$.
Ignoring the contribution of the glueball component, their flavour content,
obtained from the reactions
$p\bar p\to\pi^0\pi^0$, $\eta\eta$, $\eta\eta'$, is

\begin{eqnarray}
f_2(1920) &=& \cos\varphi_{n=3}n\bar
n+\sin\varphi_{n=3}s\bar s, \nonumber\\ f_2(2120) &=&
-\sin\varphi_{n=3}n\bar n+\cos\varphi_{n=3}s\bar s, \nonumber\\ &&
\varphi_{n=3}\ =\ 0\pm5^\circ.
\label{13}
\end{eqnarray}

\item The next, predominantly $^3P_2$ states with $n=4$ are
$f_2(2240)$ and $f_2(2410)$ \cite{LL}. (By mistake, in \cite{PDG}
the resonance $f_2(2240)$ \cite{Ani} is listed as $f_2(2300)$,
while $f_2(2410)$ \cite{LL} is denoted as $f_2(2340)$). Their
flavour content at $W=0$ is determined as

\begin{eqnarray}
f_2(2240) &=&
\cos\varphi_{n=4}n\bar n+\sin\varphi_{n=4}s\bar s, \nonumber\\
f_2(2410) &=& -\sin\varphi_{n=4}n\bar n+\cos\varphi_{n=4}s\bar s,
\nonumber\\
&& \varphi_{n=4}\ =\ 5\pm11^\circ.
\label{14}
\end{eqnarray}

\item $f_2(2020)$ and $f_2(2340)$ \cite{LL} belong to the basic $F$-wave
nonet $(n=1)$ (in \cite{PDG} the $f_2(2020)$ \cite{Ani} is
denoted as $f_2(2000)$ and is put in the section "Other light mesons",
while $f_2(2340)$ \cite{LL} is denoted as $f_2(2300)$). The flavour
content of the $1\,^3F_2$ mesons is

\begin{eqnarray}
f_2(2020) &=& \cos\varphi_{n(F)=1}n\bar n+\sin\varphi_{n(F)=1}s\bar s,
\nonumber\\
f_2(2340) &=& -\sin\varphi_{n(F)=1}n\bar n+\cos\varphi_{n(F)=1}s\bar s,
\nonumber\\
&& \varphi_{n(F)=1}\ =\ 5\pm8^\circ.
\label{15}
\end{eqnarray}

\item The resonance $f_2(2300)$ \cite{Ani} has a dominant $F$-wave
quark-antiquark component; its flavour content for $W=0$ is
defined as

\begin{equation}
\label{16}
f_2(2300)=\cos\varphi_{n(F)=2}n\bar n+\sin\varphi_{n(F)=2}s\bar s,
\quad \varphi_{n(F)=2}=-8^\circ\pm12^\circ.
\end{equation}
A partner of $f_2(2300)$ in the $2\,^3F_2$ nonet has to be a $f_2$-resonance
with a mass $M\simeq2570\,$MeV.

\end{enumerate}

\begin{figure}[h]
\centerline{\epsfig{file=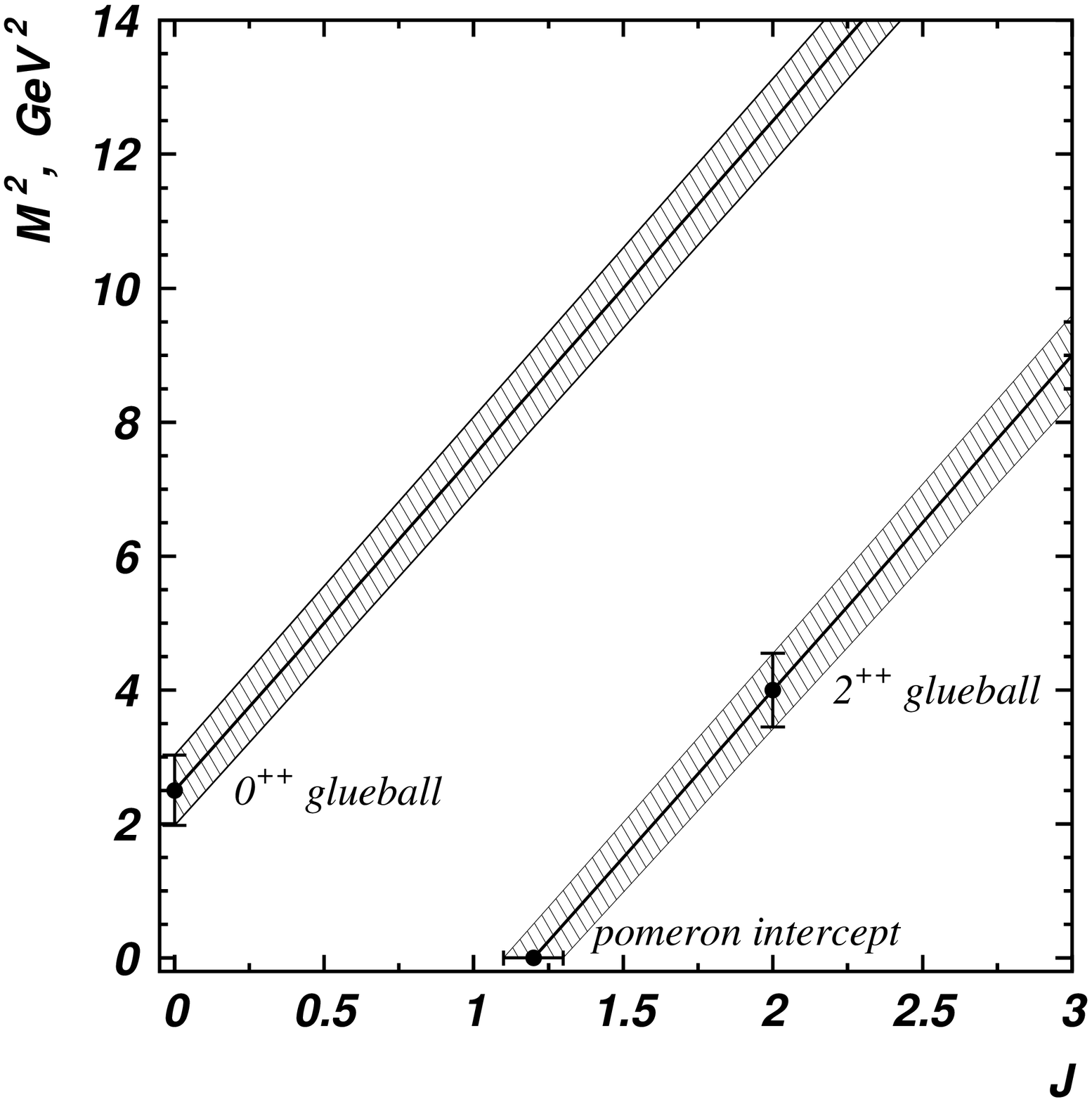,width=10cm}}
\caption{Glueball states on the Pomeron trajectories.}
\end{figure}

\section{Conclusion}

The broad $f_2(2000)$ state is the descendant of the lowest tensor
glueball. This statement is favoured by estimates of parameters of
the Pomeron trajectory (e.g., see \cite{book}, Chapter 5.4, and
references therein), according to which $M_{2^{++}glueball}\simeq
1.7-2.5$ GeV. Lattice calculations result in a similar value,
namely, 2.2--2.4 GeV \cite{lattice}. The corresponding coupling
constants $f_2(2000)\to\pi^0\pi^0, \eta\eta, \eta\eta'$ satisfy
the relations for the glueball, eq.(\ref{4.21}), with
$\lambda\simeq0.5-0.7$. The admixture of the quarkonium component
$(q\bar q)_{glueball}$ in $f_2(2000)$ cannot be determined by the
ratios of the coupling constants between the hadronic channels; to
define it, $f_2(2000)$ has to be observed in
$\gamma\gamma$-collisions. The value of $(q\bar q)_{glueball}$ in
$f_2(2000)$ may be rather large: the rules of $1/N$-expansion give
a value of the order of $N_f/N_c$. It is, probably, just the
largeness of the quark-antiquark component in $f_2(2000)$ which
results in its suppressed production in the radiative $J/\psi$
decays (see discussion in \cite{Bugg}).

We have now two observed glueballs, a scalar $f_0(1200-1600)$
\cite{APS-PL,AAS-PL} (see also \cite{book,ufn04}) and a tensor
one, $f_2(2000)$. It is illustrative to present the situation with
$0^{++},2^{++}$ glueballs on the $(J,M^2)$-plane, we demonstrate
this in Fig. 18. According to various estimates, the leading
Pomeron trajectory has an intercept at $\alpha (0)\simeq 1.10 -
1.30$ (see, for example, \cite{kaid,land,dakhno}). Assuming that
the Pomeron trajectory has a linear behaviour, which does not
contradict experimental data,
$\alpha_P(M^2)=\alpha_P(0)+\alpha'_P(0)M^2$, we have for the slope
$\alpha'_P(0)=0.20\pm 0.05$. The scalar glueball $f_0(1200-1600)$
is located on the daughter trajectory which predicts the second
tensor glueball at $M\simeq 3.45$ GeV. If the Pomeron trajectories in
$(n,M^2)$ plane are linear, similar to the $q\bar q$-trajectories, then
the next scalar glueball (radial exitation of $gg$ gluonium) should be
at $M\simeq 3.2 GeV$.

Observed glueball states have transformed into broad resonances
owing to the accumulation of widths of their neighbours. The
existence of a low-lying pseudoscalar glueball is also expected. It
is natural to assume that it has also transformed into a broad
resonance. Consequently, the question is, where to look for this
broad $0^{-+}$ state. There are two regions in which we can suspect
the existence of a pseudoscalar glueball: in the region of 1700~MeV
or much higher, at $\sim2300\,$MeV, see the discussion in
\cite{Bugg}\ (Section 10.5). In \cite{Faddeev} it is suggested that
the lowest scalar and pseudoscalar glueballs must have roughly equal
masses. If so, a $0^{-+}$ glueball has to occur in the 1700~MeV
region.

The authors are grateful to D.V.~Bugg, L.D.~Faddeev and S.S.~Gershtein
for stimulating discussions. The paper was supported by the grant
No. 04-02-17091 of the RFFI.

\end{document}